\documentclass[runningheads]{llncs}
\usepackage{graphicx}
\usepackage{microtype}
\usepackage{graphicx}
\usepackage{subfigure}
\usepackage{booktabs}

\usepackage{amssymb,amsfonts,amsmath}
\usepackage{xcolor}
\usepackage{xspace}
\usepackage{caption}
\usepackage{algorithm}
\usepackage{wrapfig,epsfig}
\usepackage{multirow}
\usepackage{setspace}
\usepackage{enumitem}

\usepackage{algorithm}
\usepackage{algorithmic}

\DeclareMathOperator*{\argmax}{arg\,max}

\newcommand{\fedavg}{\texttt{FedAvg}\xspace}
\newcommand{\fedpp}{\texttt{FedP2P}\xspace}

\begin{document}
%
%%\title{Contribution Title\thanks{Supported by organization x.}}
\title{Efficient and Less Centralized\\Federated Learning}
\author{Li Chou*\inst{1} \and
Zichang Liu*\inst{1} \and
Zhuang Wang\inst{1}  \and
Anshumali Shrivastava\inst{1}
}

% %\authorrunning{F. Author et al.}

\institute{Rice University, Houston TX, USA 
\email{\{lchou,zl71,zw50,anshumali\}@rice.edu}}

\maketitle              % typeset the header of the contribution
\begin{abstract}
With the rapid growth in mobile computing, massive amounts of data and computing resources are now located at the edge. To this end, Federated learning (FL) is becoming a widely adopted distributed machine learning (ML) paradigm, which aims to harness this expanding skewed data locally in order to develop rich and informative models. In centralized FL, a collection of devices collaboratively solve a ML task under the coordination of a central server. However, existing FL frameworks make an over-simplistic assumption about network connectivity and ignore the communication bandwidth of the different links in the network. In this paper, we present and study a novel FL algorithm, in which devices mostly collaborate with other devices in a pairwise manner. Our nonparametric approach is able to exploit network topology to reduce communication bottlenecks. We evaluate our approach on various FL benchmarks and demonstrate that our method achieves $10\times$ better communication efficiency and around 8\% increase in accuracy compared to the centralized approach.

%%The abstract should briefly summarize the contents of the paper in
%%150--250 words.

%%\keywords{First keyword  \and Second keyword \and Another keyword.}
\end{abstract}

\section{Introduction}
The rapid growth in mobile computing on edge devices, such as smartphones and tablets, has led to a significant increase in the availability of distributed computing resources and data sources. These devices are equipped with ever more powerful sensors, higher computing power, and storage capability, which is contributing to the next wave of massive data in a decentralized manner. To this end, federated learning (FL) has emerged as a promising distributed machine learning (ML) paradigm to leverage this expanding computing and data regime in order to develop information-rich models for various tasks. At a high-level, in FL, a collection or federation of devices collaboratively solve a ML problem (i.e., learn a global model) under the coordination of a centralized server. The crucial aspect is to accomplish the task while maintaining the data locally on the device. With powerful computing resources (e.g., cloud servers), FL can scale to millions of mobile devices~\cite{fl_survey}. However, with the continual increase in edge devices, one of the vital challenges for FL is communication efficiency \cite{bonawitz&etal_sysml2019,fl_survey,li_fedprox}.

Unlike classical distributed ML, where a known architecture is assumed, the structure for FL is highly heterogeneous in terms of computing resources, data, and network connections. Devices are equipped with different hardware, and are located in dynamic and diverse environments. In these environments, network connections can have a higher failure rate on top of varying communication and connection patterns. For example, 5\% and more of the devices participating in the single round of training may fail to complete training or completely drop out of communication~\cite{fl_survey}. The device network topology may evolve, which can be useful information for communication efficiency. In FL, the central server has the critical role of aggregating and distributing model parameters in a back and forth manner (i.e., rounds of communication) with the devices to build and maintain a global model. One natural solution is to have powerful central servers. However, 
this setup comes with high costs and are only affordable for large corporations~\cite{fl_survey}. Moreover, overly relying on a central server can suffer from single point failure~\cite{vanhaesebrouck&etal_aistats17} and communication bottleneck.

\begin{figure*}
	\centering
	\subfigure[]{
		\label{fig:fedavg-graph}
		\raisebox{5mm}{
		\includegraphics[trim = 60 0 60 0, clip,scale=.36]{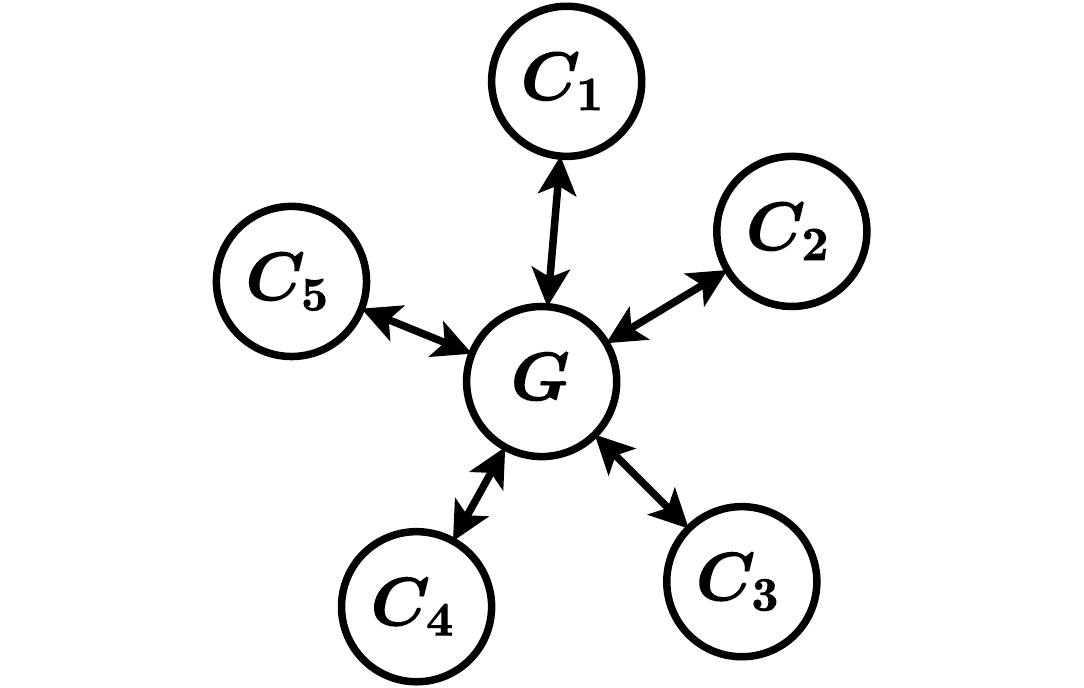}
		}
	}\hspace*{1cm}		
	\subfigure[]{
		\label{fig:fedavg-communication}
		\includegraphics[trim = 30 0 50 0, clip,scale=.36]{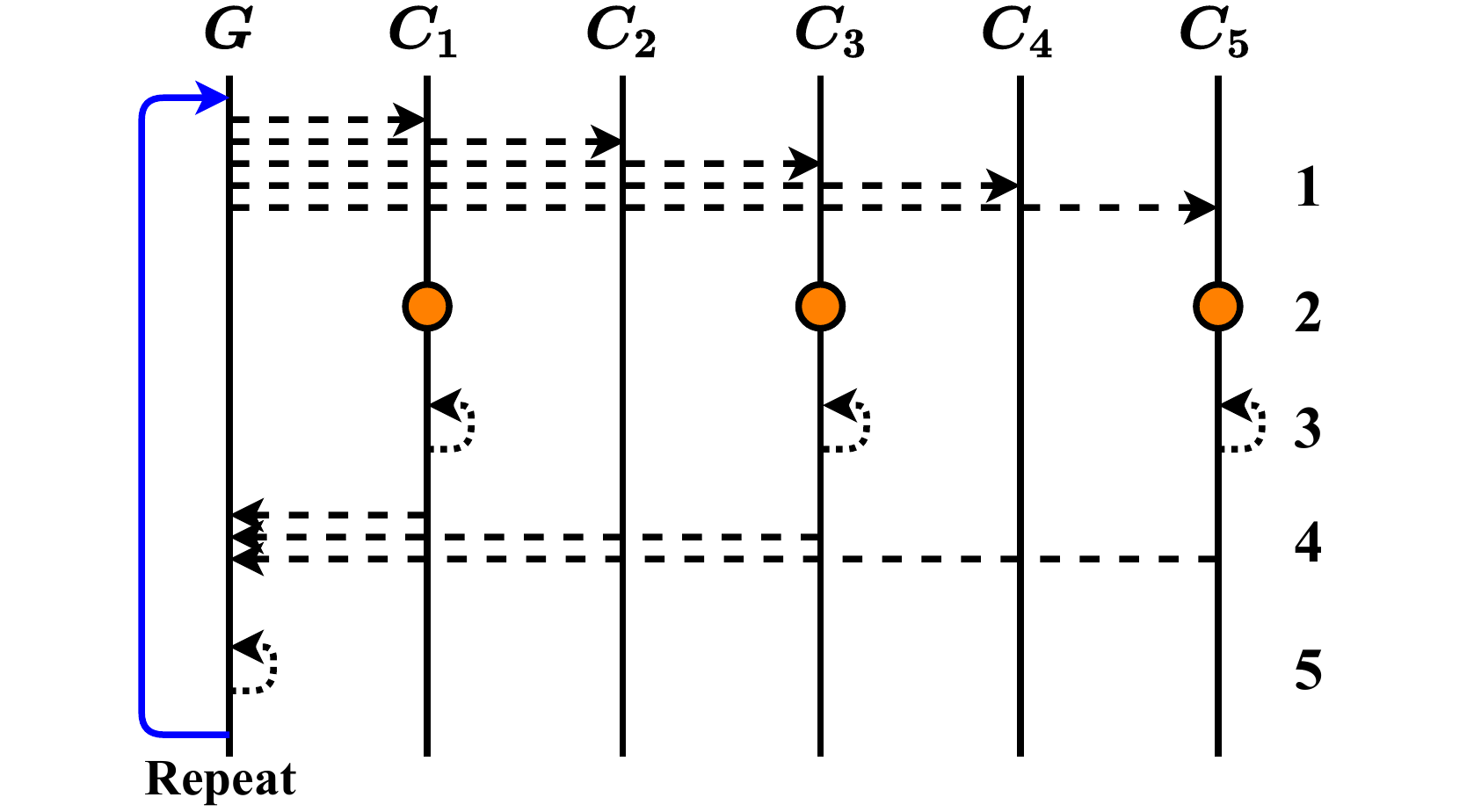}
	}\\[20pt]
	\subfigure[]{
		\label{fig:fedp2p-graph}
		\includegraphics[trim = 40 0 40 0, clip,scale=.36]{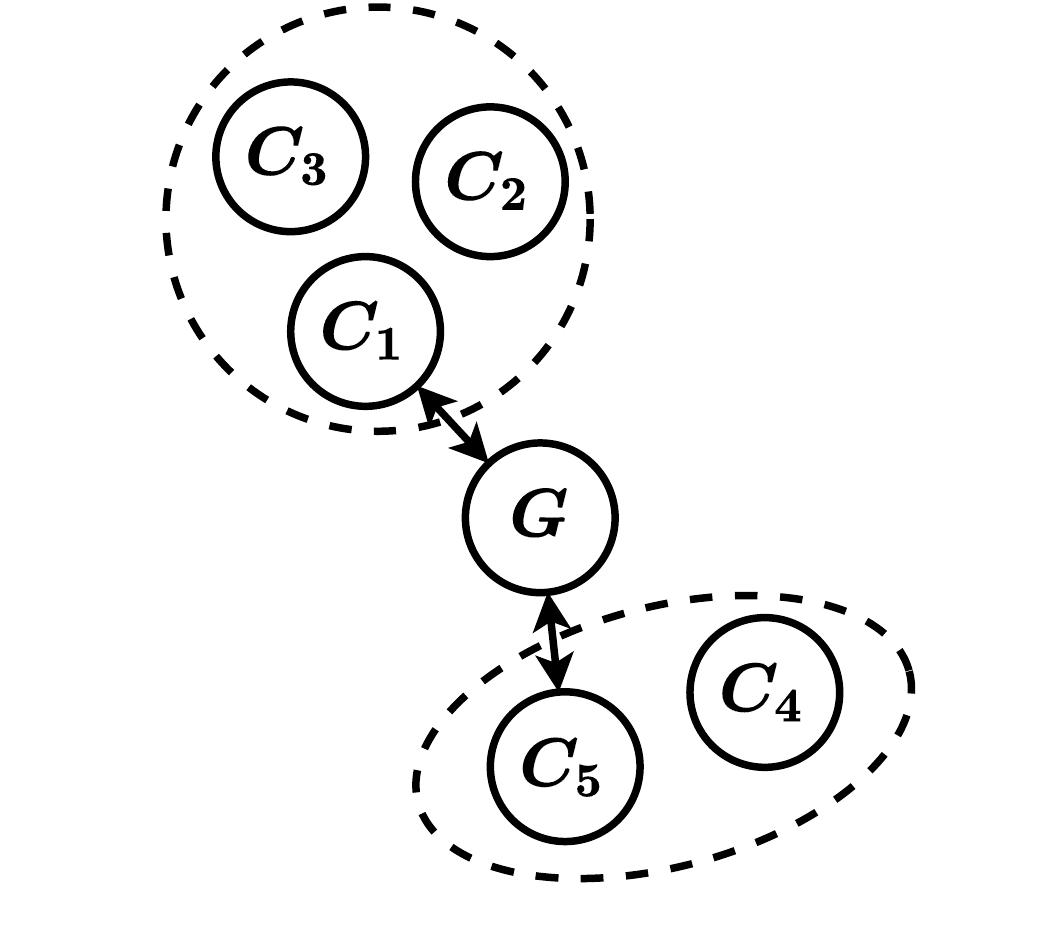}
	}\hspace*{1cm}
	\subfigure[]{
		\label{fig:fedp2p-communication}
		\includegraphics[trim = 30 0 40 0, clip,scale=.36]{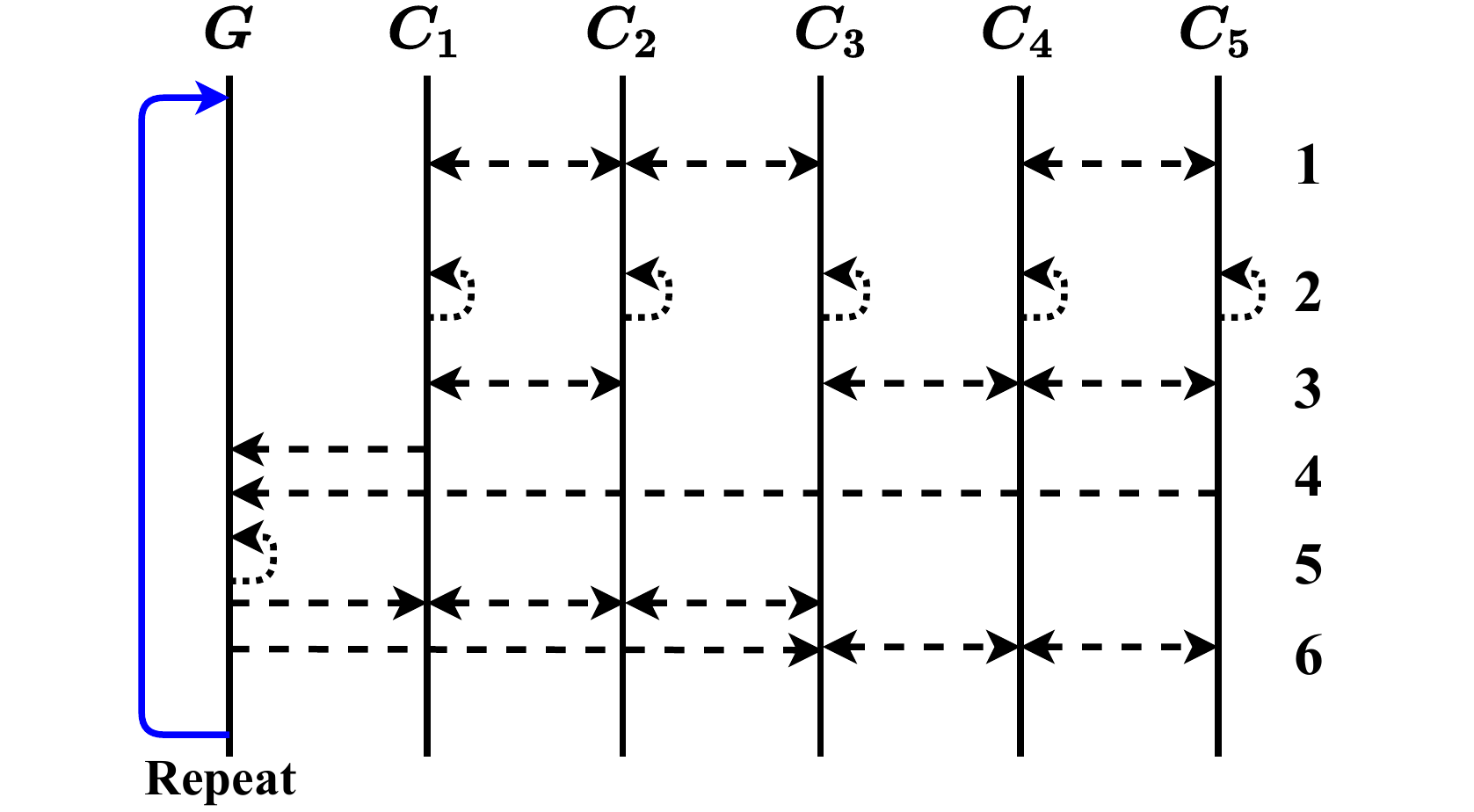}
	}
%	\vspace*{-3mm}
	\caption{(a) Structure for centralized FL framework, in which central server directly communicates with all devices for model distribution and aggregation (b) Communication flow for centralized FL (\fedavg). The steps are 1: server $G$ sends global model to clients, 2: $G$ designates client devices for training ($C_1,C_3,C_5)$, 3: selected client devices train on local data, 4: devices send parameters of the trained model back to $G$, 5: $G$ aggregates the trained parameters, and the process repeats. (b) Structure for our proposal (\fedpp). Dash circle represents local P2P network, in which devices perform pairwise communication. The central server only communicates with a small number of devices, one device within each local P2P network here. (d) Communication flow for \fedpp. The steps are: 1: form P2P network, 2: all devices train in parallel on local data, 3: devices can aggregate/synchronize models via Allreduce, 4: $G$ receives aggregated parameter from each partition (not from all devices), 5: $G$ aggregates the parameters from the partitions to get global parameter, and 6: $G$ sends global parameter to the P2P networks (not to all devices).}
	%\vspace*{-2mm}
\end{figure*}

 An alternative to the client-server architecture is peer-to-peer (P2P) networking. P2P dramatically reduces the communication bottleneck by allowing devices to communicate with one another.  Given this insight, we propose a FL framework, which we dub as \fedpp, that leverages and incorporates the attributes of a P2P setting. \fedpp significantly reduces the role and communication requirements of the central server. Additionally, P2P naturally utilizes network topology to better structure device connections. It is widely accepted that communication delays increase with the node degrees and spectral gap of the network graph~\cite{fl_survey}. Explicit consideration of network topology increases communication efficiency such as wall-clock time per iteration. Empirically, we show \fedpp outperforms the established FL framework, \fedavg~\cite{mcmahan_fedavg} on a suite of FL benchmark datasets on both computer vision and language tasks, in addition to two synthetic data, involving three different model architectures. With the same number of communication rounds with a central server, \fedpp achieves an 8\% increase in accuracy. With the same number of devices participating in the training, \fedpp can achieve $10\times$ speed up in communication time.

\section{Preliminary and Related Work}

FL is a distributed ML paradigm, where data reside on multiple devices, and under the coordination of a centralized server, the devices collaboratively solve an ML problem~\cite{fl_survey}. Moreover, and importantly, data are not shared among devices, but instead, model aggregation updates, via communication between server and devices, are used to optimize the ML objective. Naturally, what we can expect as corollaries, and as key challenges, from this setting are: a massive number of devices, highly unbalanced and skewed data distribution (i.e., not identically and independently distributed, and limited network connectivity (e.g., slow connections and frequent disconnections). In FL, the optimization problem is defined as
\begin{equation}
\min_{\theta} f(\theta) \quad\quad \text{where} \quad\quad f(\theta) \triangleq \sum_{i=1}^{N}p_{i} F_{i}(\theta).
\label{eq:fl_opt}
\end{equation}
$N$ is the number of devices, $p_{i} \in [0,1]$, and $\sum_{i=1}^{N}p_{i}=1$.  For supervised classification, we typically have $F_{i}(\theta) = \ell(X_{i},Y_{i};\theta)$, where $(X_{i},Y_{i})$ are the data and $\theta$ are the model parameters. $\ell$ is some chosen loss function for client devices $C_i$. We assume data is partitioned across $N$ client devices, which give rise to potentially different data distributions. Namely, given data $\mathcal{D}$ partitioned into samples $D_1, \ldots, D_N$ such that $D_i$ corresponds to $C_i$, the local expected empirical loss is $F_i(\theta) = \mathbb{E}_{X_i \sim D_i}[f_i(X_i; \theta)]$ where $p_i{=}|D_i|/|\mathcal{D}|$. $D_i$ varies from device to device and is assumed to be non-IID. The federated averaging (\fedavg) algorithm~\cite{mcmahan_fedavg} was first introduced to optimize the aforementioned problem,  which we describe next.

\subsection{Centralized Federated Learning: Federated Averaging}

\fedavg is a straightforward algorithm to optimize Eq.~(\ref{eq:fl_opt}). The details are shown in Algo.~\ref{alg:fedavg} and Fig.~\ref{fig:fedavg-communication}. The global objective, orchestrated by a central server $G$, is optimized locally via the $F_{i}$ functions applied to the local data on the devices. Let $\mathcal{C}=\{C_1,\ldots,C_N\}$ be the collection of client devices that are willing and available to participate in training a global model for a specific machine learning task. At each training round $t$, the central server randomly selects a subset $Z \subset \mathcal{C}$ devices to participate in round $t$ of training.  $G$ sends to each selected device a complete copy of the global model parameters $\theta_G^{t}$. Each device trains $F_{i}$, via stochastic gradient descent (SGD), initialized with $\theta_G^{t}$ using its own training data $D_i$ for a fixed number of epochs $E$, batch size $O$, and according to a fixed learning rate $\eta$. Subsequently, the client devices sends the updated parameters $\theta_{C_i}^{t+1}$ back to $G$, where $G$ performs a model synchronization operation. This process is repeated for $T$ rounds, or until a designated stopping criterion (e.g., model convergence). Unlike traditional distributed machine learning, \fedavg propose to average model parameters (i.e., Aggregate$(\cdot)$ operation in Algo.~\ref{alg:fedavg}) instead of aggregating gradients to dramatically reduce communication cost. \fedavg started an active research field on centralized federated learning, which assumes that all communications occur directly with the central server. 

\begin{algorithm}[t]
    \caption{Federated Averaging (\fedavg)}
    \label{alg:fedavg}
	\begin{algorithmic}
		\STATE {\bfseries Input:} $T, |Z|, \eta, O, E, \theta^0_G$
		\FOR{$t = 0 \ $ to $ \ T{-}1$}		
			\STATE Server $G$ samples subset of client devices $C_i \in Z$	
				\STATE and sends $\theta^t_G$ to sampled devices: $\theta^t_{C_i} \leftarrow \theta^t_G$
			\FOR{{\bf each} $C_i \in Z$ {\bf in parallel}}
				\STATE {\footnotesize \texttt{//Device trains on local data with}}
				\STATE {\footnotesize \texttt{//step size}}$ \ \eta, \ $ {\footnotesize \texttt{batch size}}$ \ O, \ ${\footnotesize \texttt{epochs}}$ \ E$
				\STATE $\theta_{C_i}^{t+1} \leftarrow \min F_i(\theta_{C_i}^{t})$
			\ENDFOR
		\ENDFOR
		\STATE $\theta_{G}^{t+1} \leftarrow$ Aggregate $\left(\theta_{C_{1}}^{t+1}, \forall C_i \in Z\right)$		
	\end{algorithmic}
\end{algorithm}

\subsection{Multi-Model Centralized Federated Learning}

Here we focus on related FL settings, where the goal is to learn multiple models using the idea of clustering. These methods mostly utilize the information from model parameters for device clustering. All these methods concentrate on optimization while ignoring practical communication constraints.  For example, \cite{briggs&etal} uses a hierarchical $k$-means clustering technique based on similarities between the local and global updates. At every round, all devices are clustered according to the input parameter $k$, which makes this technique not practical in a real-world FL setting. \cite{ghosh&etal_neurips20} focuses on device clustering. They assume $k$ different data distributions, and the server maintains $k$ different global models. A subset of devices is sampled at each round, and all $k$ models are sent to each of the devices. The sampled devices then optimize the local objective, using each of the $k$ models, and selects the model with the lowest loss. The updated model with the lowest loss, which also corresponds to the device's cluster identity, is sent back to the server for aggregation based on the clusters. The process is repeated for a given number of rounds. This method can be interpreted as a case of $k$-means with subsampling. The need to communicate $k$ models to each device creates an information communication bottleneck. We emphasize that our goal is not to solve a clustering problem.

\subsection{Decentralized Federated Learning}

Decentralized FL is also an active area of research, and we focus on the corresponding FL setting that involves P2P communication in this section. FL utilizing P2P communication has been previous proposed by ~\cite{lalitha&etal_arxiv19}. However, a graph structure is imposed, and communication is based on the graph structure and limited to one-hop neighbors. Also, graph topology can change over time. The algorithm requires potentially complex Bayesian optimization. In addition, the experiments are limited to 2 nodes. 

P2P ML, where the goal is to learn personalized models, as opposed to a single global model, has been proposed by in~\cite{vanhaesebrouck&etal_aistats17} and under strong privacy requirements by~\cite{bellet&etal_aistats18}. The gossip protocol is a P2P communication procedure based on how epidemics spread~\cite{demers&etal_gossip}. Gossip algorithms are distributed asynchronous algorithms with applications to sensors, P2P, and ad-hoc networks. It has been used to study averaging as an instance of the distributed problem~\cite{boyd&etal_gossip}, and successfully applied in the area of decentralized optimization~\cite{colin&etal_icml16,koloskova&etal}. Also, several works focus on decentralized FL from the perspective of optimization, decentralized SGD \cite{lian&etal_neurips17,tang&etal_icml18,local_sgd}.

\subsection{Efficient Pairwise Communication}

Allreduce \cite{ring-allreduce2009,MPIoptimization} is a collective operation that reduces the target tensors in all processes to a single tensor with a specified operator (e.g., sum or average) and broadcasts the result back to all processes.
It is a decentralized operation and only involves P2P communication for the reduction (e.g., model synchronization).
Because Allreduce is a bandwidth-optimal communication primitive and is well-scalable for distributed training, it is widely adopted in distributed ML frameworks~\cite{pytorch,horovod}.

\section{Less Centralized Federated Learning}
In this section, we outline our federated learning framework, \fedpp, and provide a detailed discussion of the critical aspects of our design. With \fedpp, we exploit efficient P2P communication in conjunction with a coordinating center that does high-level model aggregation.

\subsection{Proposed Framework: \fedpp}

The structure of centralized FL follows a star graph, in which the central server directly communicates with all client devices (see Fig.~\ref{fig:fedavg-graph}). On the contrary, we aim to reorganize the connectivity structure in order to distribute both the training and communication on the edge devices by leveraging P2P communication. To this end, we form $L$ local P2P networks in which client devices perform pairwise communication within each P2P network, which we refer to as \fedpp. With \fedpp, the central server only communicates with a small number of devices, one from each local P2P network (see Fig.~\ref{fig:fedp2p-graph}).

Here, we describe the training process. The details are shown in Algo.~\ref{alg:fedp2p} and Fig.~\ref{fig:fedp2p-communication}. To better understand our framework, we follow the centralized setup by describing the whole training process in $T$ rounds. We describe the process as three phases for each round as follows.
\begin{enumerate}
%\begin{enumerate}[leftmargin=*, itemsep=0em, topsep=-2px]
%\setlength\itemsep{0em}
    \item \textbf{Form Local P2P Network}: At the start of each round $t$, the central server randomly partitions $N$ devices into $L$ local P2P networks $Z_{1}, \ldots, Z_{L}$. The central servers distribute the global model $\theta^{t-1}$ from the previous round to each P2P network. Note that the central server only communicates with one or a few devices from each P2P network for the global model distribution. In practice, $L$ is not a tuning parameter, but can be precisely calculated to minimize communication cost given devices bandwidth and the desired number of total participating devices in each round. We provide more information on the choice of $L$ in Section \ref{sec:comm}.
    \item \textbf{P2P Synchronization}: Within a local P2P network $Z_l$, part of or all devices train the local P2P network model $\theta_{Z_l}^t$ using its own data in parallel. Once the devices finish training, the model is locally synchronized within $Z_l$. This is done via an Aggregate$(\cdot)$ operation, which we define as $\theta_{Z_{l}^{t+1}} \leftarrow  \sum_{C_i{\in}Z} \gamma_i \theta_{C_i}$, where $\gamma_i = |D_i|/\sum_{i}|D_i|$. This process can be conducted one or more times, and we can efficiently accomplish a single P2P network model synchronization using the Allreduce approach. Note that all training and communication inside each local P2P network are conducted independently and in parallel.
    \item \textbf{Global Synchronization}: The central server $G$ globally aggregates the updated models,  $\theta_{Z_l}^{t+1}$ for $l = 1, \dots, L$, from every local P2P network, and perform model averaging over $L$ models. Namely, $\theta_G^{t+1} \leftarrow L^{-1}\sum_{l=1}^L \theta_{Z_l}^{t+1}$. Since each local P2P network is already locally synchronized, $G$ gathers models from one device for each P2P network.
\end{enumerate}

\begin{algorithm}[t]
	\caption{Federated Peer-to-Peer (\texttt{FedP2P})}
	\label{alg:fedp2p}
	\begin{algorithmic}
		\STATE {\bfseries Input:} $T, L, Q, \eta, O, E$
		\FOR{$t = 0$ to $T{-}1$ {\bf in parallel}}   
			\STATE $Z_{1},Z_{2},\ldots,Z_{L}\quad\quad$ {\footnotesize \texttt{//Form P2P networks}}
			\FOR{$l = 1$ to $L$ {\bf in parallel}}
				\FOR{$C_i \in Z \subseteq Z_l$ s.t. $|Z|=Q$ {\bf in parallel}}
					\STATE {\footnotesize \texttt{//Device trains on local data with}}
					\STATE {\footnotesize \texttt{//step size}}$ \ \eta, \ $ {\footnotesize \texttt{batch size}}$ \ O, \ ${\footnotesize \texttt{epochs}}$ \ E$
					\STATE $\theta_{C_{i}}^{t+1} \leftarrow \min F_{i}(\theta_{C_{i}}^t)$
				\ENDFOR
				\STATE $\theta_{Z_l}^{t+1} \leftarrow$ Aggregate $\left(\theta_{C_{i}}^{t+1}, \forall C_i \in Z\right)${\footnotesize \texttt{//Allreduce}}
			\ENDFOR
			\STATE $\theta_{G}^{t+1} \leftarrow $ Aggregate $\left(\theta_{Z_1}^{t+1},\ldots,\theta_{Z_L}^{t+1}\right)$
		\ENDFOR
	\end{algorithmic}
\end{algorithm}

In order to obtain the global model among all P2P networks, our proposed framework also utilizes a central coordinator for global model synchronization. However, note that the communication and workload for the central coordinator in \fedpp are significantly reduced. Assume both \fedpp and \fedavg utilizes $P$ participating devices in a single round of training. The \fedavg central server communicates with $P$ devices and perform model synchronization among $P$ models. In contrast, the \fedpp central server only communicates with $K$ devices and perform model synchronization among $K$ cluster models where $P \gg K$. From another perspective,  we assume the central server has a fixed bandwidth that can be devoted to coordinating federating training. \fedpp allows for many more devices to participate within each global round, enabling the global model to train on more data in a single global communication round. \fedpp distributes both computation workloads and communication burden to the edge, and we will provide a detailed comparison of communication cost in Section~\ref{sec:comm}.

Potential privacy and trust problems of \fedpp arise from two levels of cooperation: 1) the cooperation between the central server and the device within each local P2P network; and 2) the cooperation within each local P2P network. The first level is the same as standard centralized FL and we expect established solutions such as cryptographic protocols proposed by~\cite{bonawitz&etal_sysml2019} to work well for \fedpp. The second level follows decentralized FL settings. Existing secure aggregation protocols such as confidential smart contract~\cite{fl_survey} can be adopted. 

\subsection{Communication Efficiency}\label{sec:comm}

The central server bandwidth becomes the performance bottleneck in \fedavg when there are a large number of sampled devices.
\fedpp alleviates this server-device bottleneck through decentralization. Namely, the server only needs to communicate with a subset of the sampled devices, which we refer to as agents.
However, it incurs additional communication overhead inside P2P networks because the agents have to communicate with other devices. \fedpp needs to balance the trade-off between the server-agent and agent-device communication overhead.

We model and analyze the communication cost of \fedavg and \fedpp in this section. To simplify the analysis, we define $B_d$ as the bandwidth between each device-device pair for the training, $B_s$ as the total bandwidth capacity from the server to the devices (i.e., uplink bandwidth of server), and $M$ as the model size. We assume all P2P networks have the same $P$ number of devices participating in one single round of model training.

{\bf Communication Efficiency of \fedavg}: There are two main steps in the data transmission: 1) the model distribution from the server to the sampled devices has the communication time of $MP/B_s$; and 2) the model aggregation from the devices to the server has the communication time of $\alpha MP/B_s$, where $1/\alpha B_s$ is the total bandwidth capacity from the devices to the server (i.e., downlink bandwidth of server). Note that $\alpha \ge 1$ since the upload bandwidth of devices is typically smaller than their download bandwidth. Let $H_{avg}$, denoting the communication time in \fedavg, be $H_{avg} = (1+\alpha)MP/B_s$.

{\bf Communication Efficiency of \fedpp}: There are four main steps in the data transmission: 1) the model distribution from the server to the agent in each cluster has the communication time of $LM/B_s$; 2) the model distribution from the agent to the sampled devices in each P2P network has communication time of $PM/LB_d$, where $P/L$ is the number of sampled devices in each P2P network; 3) local model synchronization at the end of each local training round has the communication time of $2M/B_s$ (The exact communication time of Allreduce is $\frac{2(n-1)M}{nB_s}$, where $n$ is the number of workers); and 4) global model synchronization at the end of each global training round has the communication time of $\alpha LM/B_s$. Let $H_{p2p}$ denote the communication time in \fedpp be defined as
\begin{equation*}
    \small
    H_{p2p} = \frac{(1+\alpha)LM}{B_s} + \frac{PM}{LB_d} + \frac{2M}{B_d}.
\end{equation*}
The choice of $L$ is a way to balance the trade-off between the server-agent and agent-device communication overhead.
A small $L$ results in the low server-agent communication overhead because the server just communicates with a small number of agents, while it increases the agent-device communication time because each agent needs to broadcast the model to more devices; and vice versa.
$H_{p2p}$ reaches its minimum: $\min{H_{p2p}=\frac{2M}{B_d}(\frac{P}{L}+1)}$ when $L = A\sqrt{P}$, where $A=\sqrt{\frac{B_s}{(1+\alpha)B_d}}$ is a constant in a federated learning system.
Define $R = \frac{H_{avg}}{\min{H_{p2p}}}$ and $\gamma = \frac{B_s}{B_d}$. Then
\begin{equation}
    \small
    \label{eq:comm}
    R = \frac{(1+\alpha)P}{2\sqrt{\gamma(1+\alpha)P}+2\gamma}.
\end{equation}
$R > 1$ indicates that \fedpp has lower communication overhead than \fedavg.
Eq.~(\ref{eq:comm}) shows that the value of $R$ increases with 1) the increasing number of sampled devices; 2) the decreasing gap between the bandwidth capacities of the server and devices; and 3) the increasing gap between the server's uplink and downlink bandwidth capacities.

\subsection{Theoretical Insight}
In this section, we provide some theoretical insight into the performance of \fedpp.  We instantiate the framework of~\cite{fl_conv_noniid} (i.e., Theorems 1, 2, and 3; for spacing, we omit the details) to simplify the analysis. Similarly, let $F^*$ and $F_i^*$ be the minimal values of $F$ and $F_i$. Let $\Gamma = F^* - \sum_{i=1}^N p_i F_i^*$. The magnitude of $\Gamma$ quantifies the degree of non-IID (i.e., heterogenity) such that $\Gamma$ goes to zero as the number of non-IID samples decrease. We also apply the same assumptions on smoothness, convexity, and bounds on variance and norm of the the stochastic gradient. Notation-wise, we substitute $\theta_G\,{=}\,\mathbf{w}$, $J\,{=}\,C$, $U\,{=}\,L$ as the Lipschitz constant, and $V^2\,{=}\,G^2$ to uniformly bound the expected squared norm of stochastic gradients from the top. First, \fedpp satisfies the following base case.
\begin{corollary}
Let $\sum_{l=1}^{L}|Z_l|=N$ and the aggregation at the server $G$ be defined as $\theta_G^t \leftarrow \sum_{l=1}^L \psi_l \theta_{Z_l}^t$ such that $\psi_l = |\mathcal{D}|^{-1} \sum_{C_i{\in}Z_l} |D_i|$. Then the same bound in Theorem 1~\cite{fl_conv_noniid} holds for \textup{\fedpp}. 
\end{corollary}\label{col:col_1}
Corollary 1 is restrictive since we cannot hope to compute all $\psi_l$, and the total number of devices participating varies due to the stragglers effect in practice. However, we next analyze Theorems 2 and 3 under the \fedpp scheme at the level of  server $G$, which highlights an advantage of \fedpp. To ground the analysis, under the special case of one device P2P network, we have the following.
\begin{corollary}
Let $\sum_{l=1}^{L}|Z_l|<N$ and $|Z_l|=1, \ \forall l$. Then the same bound in Theorem 2 and 3~\cite{fl_conv_noniid} holds for \textup{\fedpp}. 
\end{corollary}\label{col:col_2}
Here we focus on the $J$ term in Theorems 2 and 3. From Theorem 2, we have $J=4 K^{-1} E^2 V^2$, where $K$ is number of devices for aggregation. However, note that with \fedpp, effectively, we have $J = 4(K\sum_l |Z_l|)^{-1} E^2 V^2$ at $G$. Therefore, $J$ is reduced by a factor of $\sum_l |Z_l|$ and thus, improves the bound. Similarly, from Theorem 3, we have $J=4 (N-1)^{-1}K^{-1}(N-K) E^2 V^2$. However, at $G$, the terms $(N-K)$ shrinks, and $K^{-1}$ grows significantly under \fedpp since $K$ increases by a factor of $\sum_l |Z_l|$, thereby also improves the bound.

\begin{figure*}[t]
    \centering
    \subfigure[MNIST]{
	\includegraphics[trim = 100 0 100 50, clip, width=0.3\linewidth]{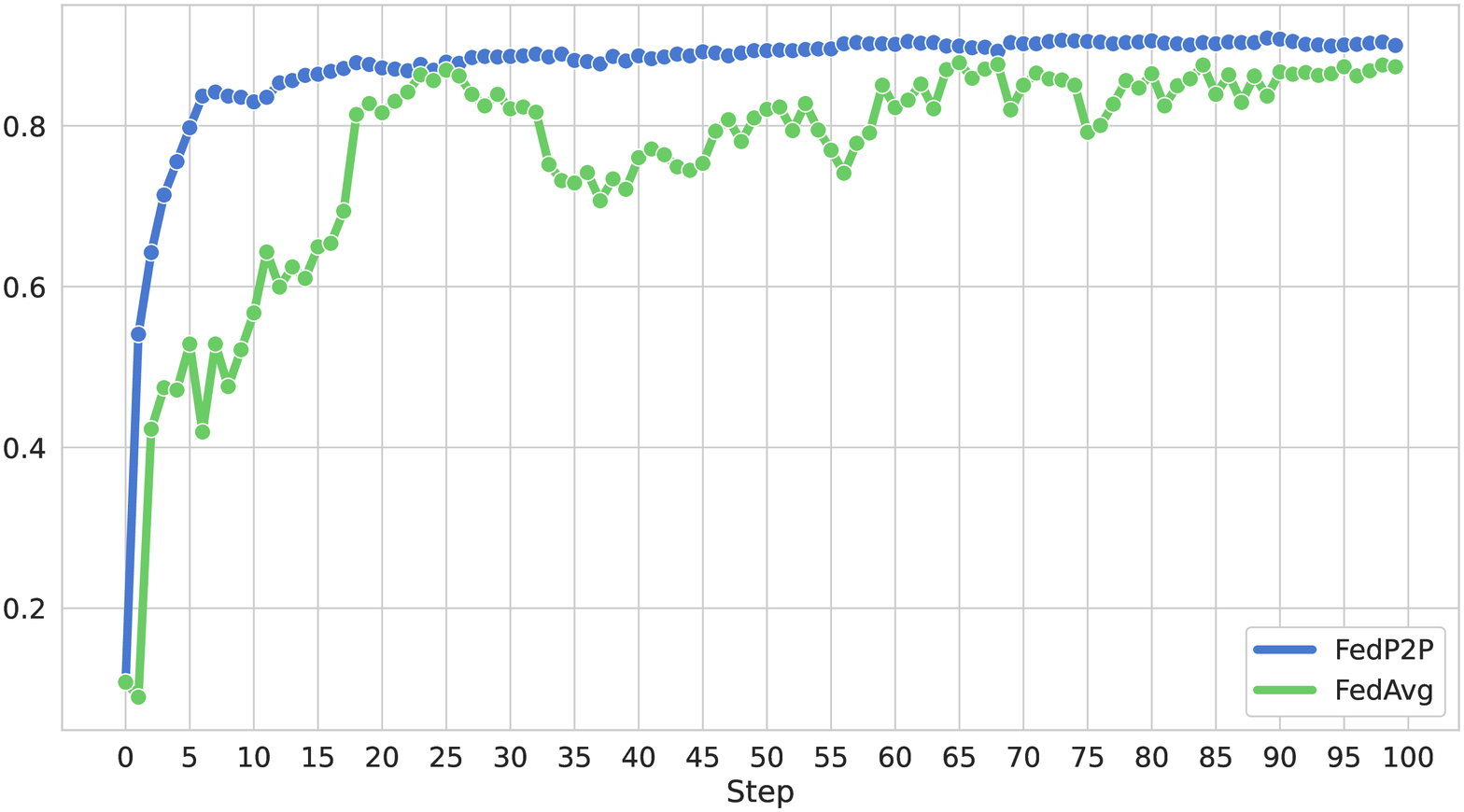}
	%\caption{MNIST}
	\label{fig:acc-1}
    }    
	\subfigure[Shakespeare]{
        \includegraphics[trim = 100 0 100 50, clip,width=0.3\linewidth]{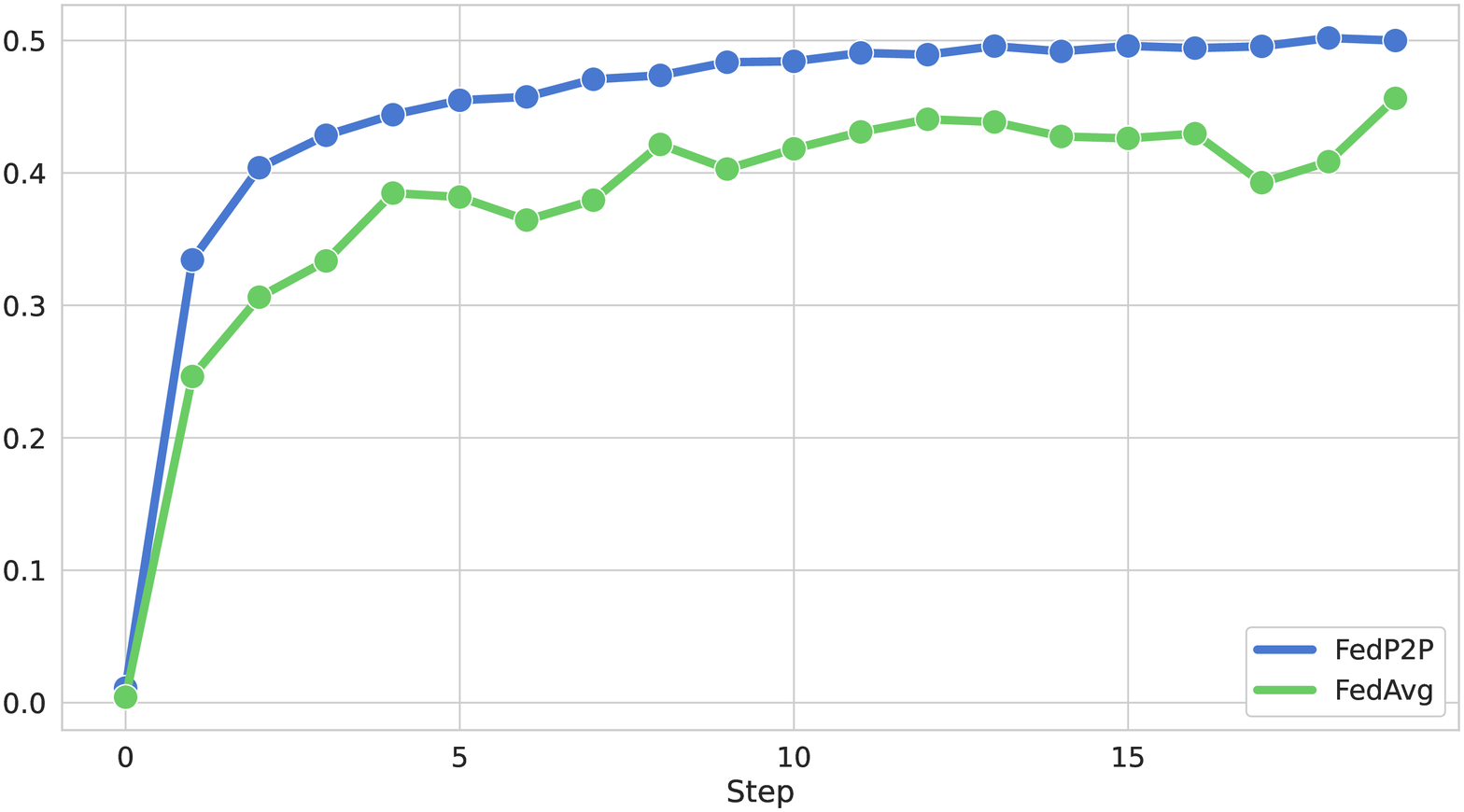}
        %\caption{SHAKESPEARE}
    \label{fig:acc-2}
    } 
	\subfigure[FEMNIST]{
        \includegraphics[trim = 100 0 100 50, clip,width=0.3\linewidth]{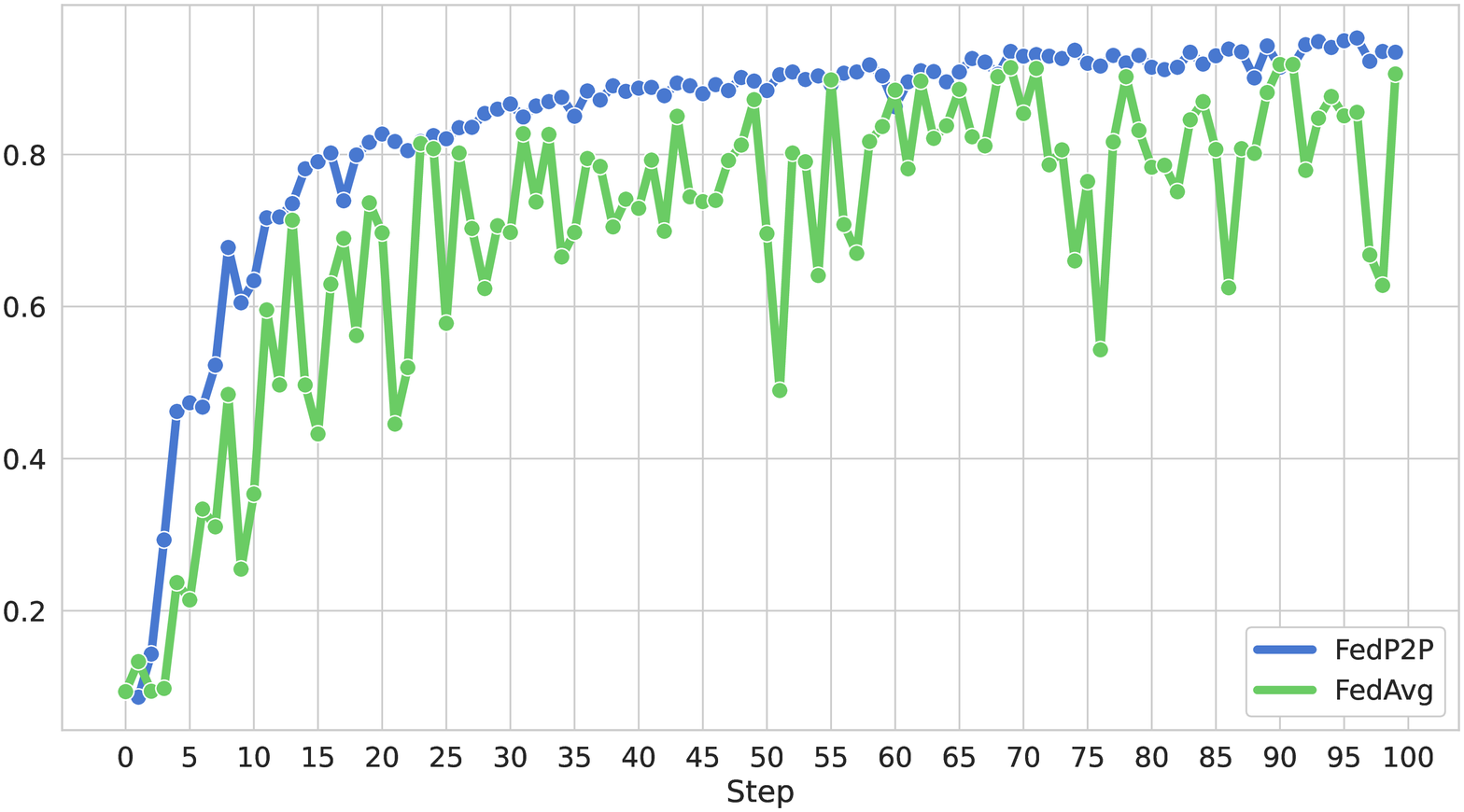}
        %\caption{FEMNIST}
        \label{fig:acc-3}
    }  
    \subfigure[SynCov]{
	\includegraphics[trim = 100 0 100 50, clip, width=0.3\linewidth]{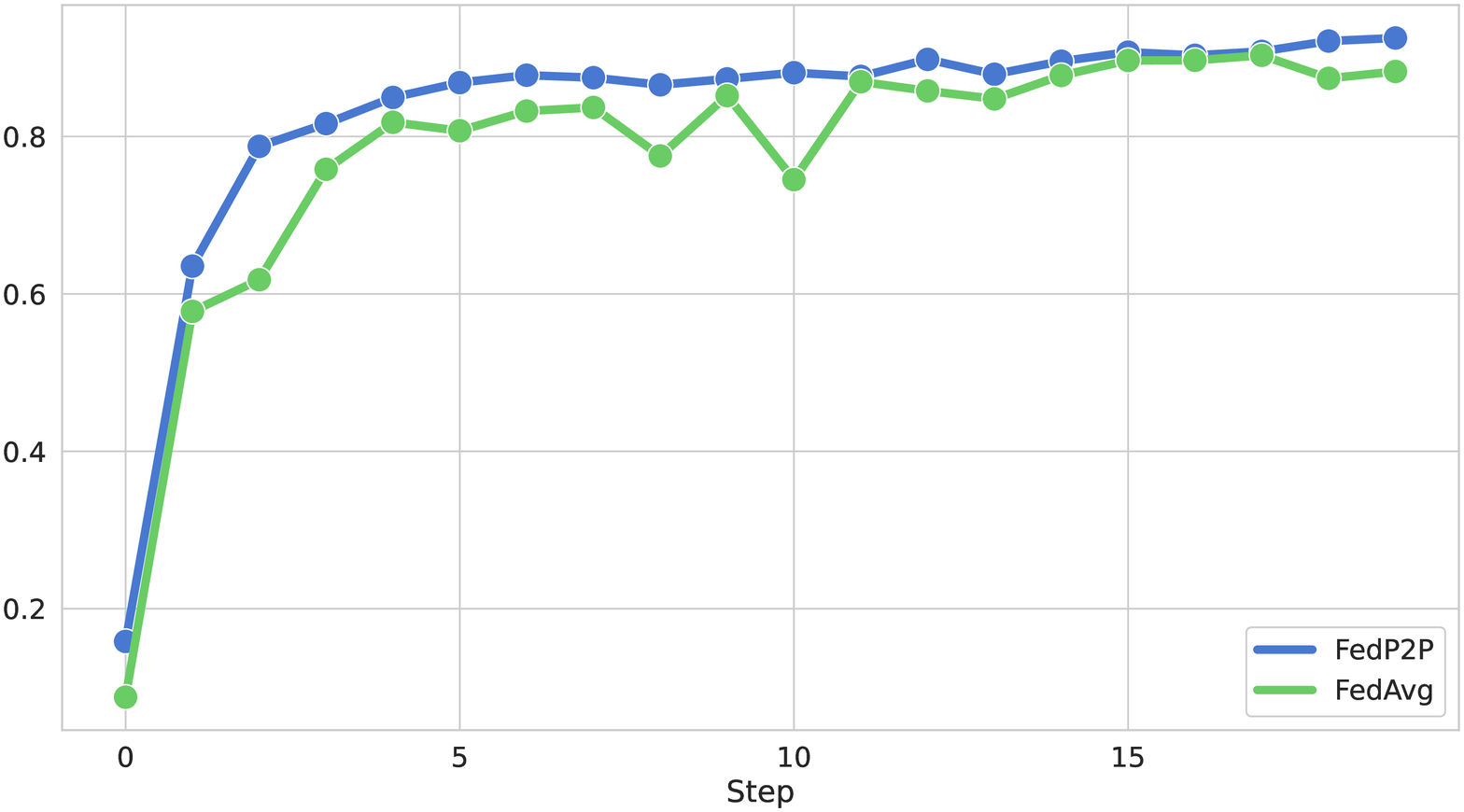}
	%\caption{SynCov}
	\label{fig:acc-4}
    }    
	\subfigure[SynLabel]{
        \includegraphics[trim = 100 0 100 50, clip,width=0.3\linewidth]{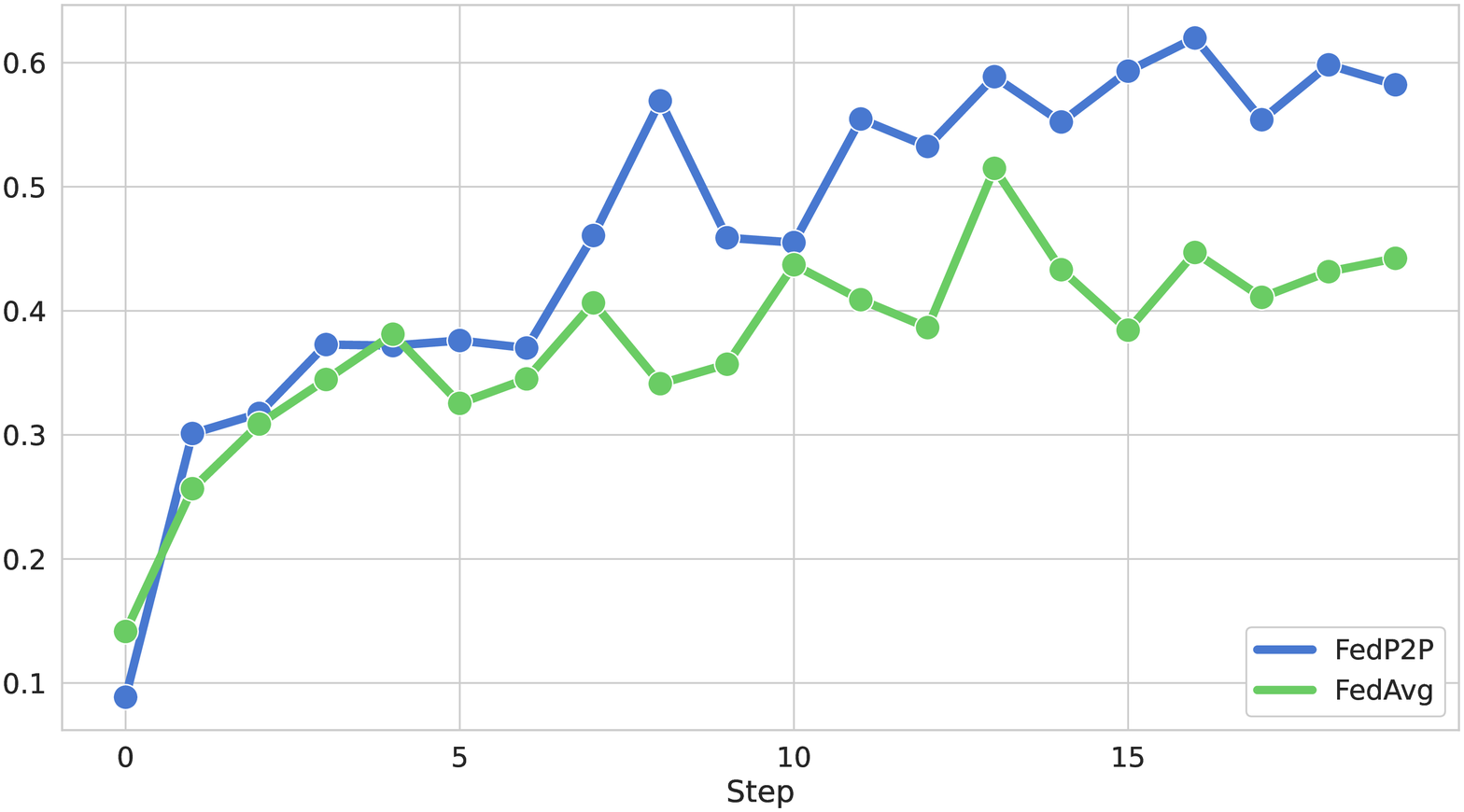}
        %\caption{SynLabel}
        \label{fig:acc-5}
    } 
	\subfigure[Training Loss (FEMNIST)]{
        \includegraphics[trim = 100 0 100 50, clip,width=0.3\linewidth]{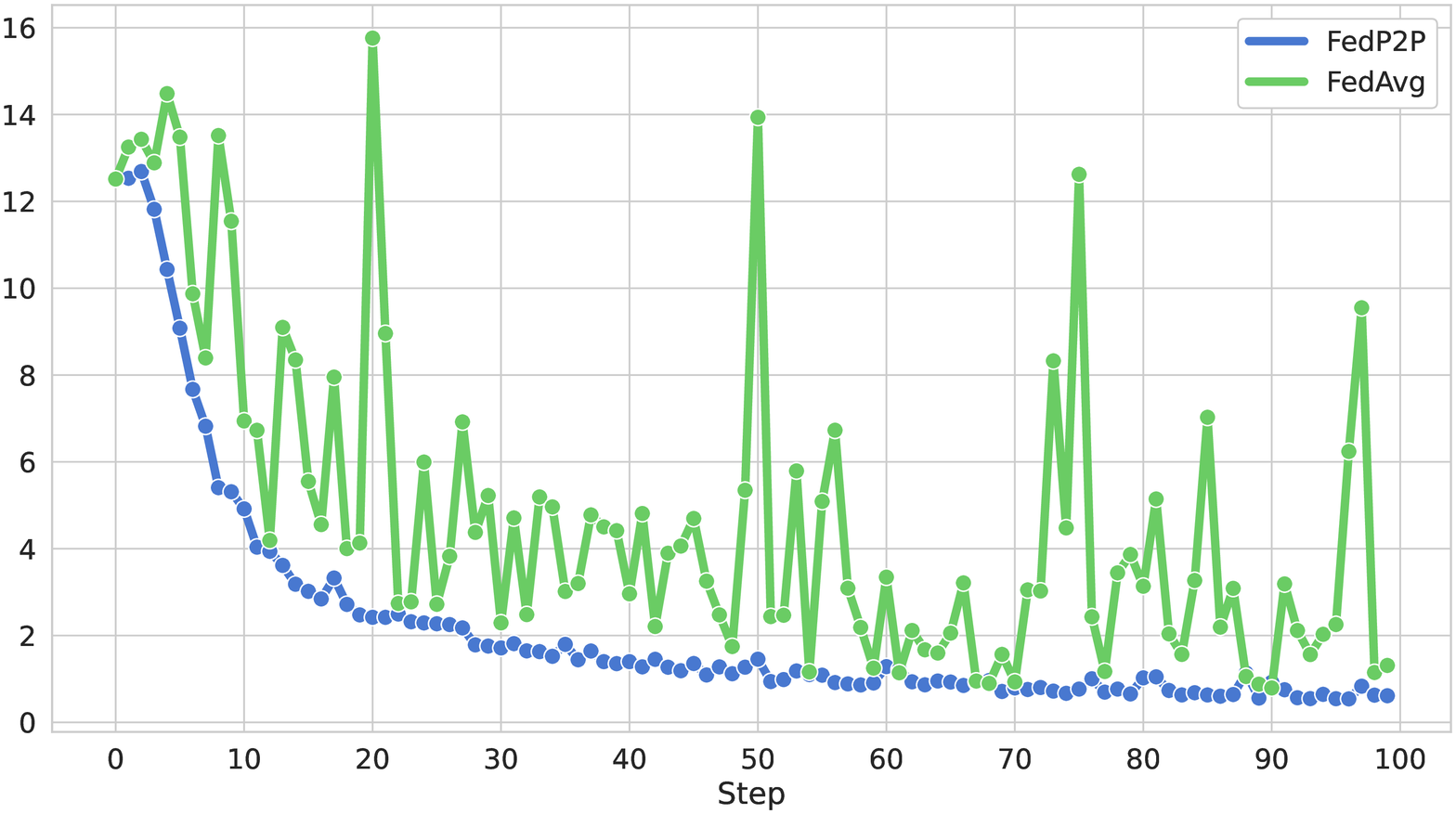}
        %\caption{Training Loss(FEMNIST)}
        \label{fig:acc-6}
    } 
%    \vspace*{-3mm}
    \caption{ \ref{fig:acc-1}, \ref{fig:acc-2}, \ref{fig:acc-3}, \ref{fig:acc-4}, and \ref{fig:acc-5} are average testing accuracy across devices as a function of global communication rounds. \fedpp shows higher and much smoother accuracy curves on all real datasets and two synthetic datasets. \ref{fig:acc-6} plots the loss during training on FEMNIST. \fedpp gives a smooth convergence curve. }
%    \vspace*{-2mm}
    \label{fig:mainacc}
\end{figure*}

\section{Experiments}

In this section, we present the experiment results that demonstrate the performance between \fedpp and \fedavg from various perspectives. Specifically, we focus on answering the following questions: 1) how does \fedpp perform comparing to centralized methods on model accuracy? 2) does \fedpp  outperform centralized federated learning in terms of communication efficiency? 3) How robust is \fedpp  handling network instability such as stragglers? 4) How does \fedpp perform with different parameters? 

\subsection{Datasets}

\textbf{Synthetic Datasets}: In FL, the data on devices are non-IID. Following the taxonomy described in~\cite{fl_survey}, non-IID is decomposed into five regimes: 1) covariance shift, 2) label probability shift, 3) same label with different features, 4) same features with different labels, and 5) quantity skew. To effectively explore and understand the various non-IID situations, we create two synthetic datasets: SynCov and SynLabel. The setup is similar to~\cite{li_fedprox}. SynCov simulates covariance shift with quantity skew, and SynLabel simulates label probability shift with quantity skew. The high-level data generation process is as follows. First, we designate $N$($=100$) number of client devices $C_i$. Next, we sample from a lognormal distribution to determine the number of data points for each $C_i$. Let $X$ be the features and $Y$ be the class labels, where feature dimension is $60$ and number of classes is $10$. Then, we sample $(X,Y)$ from the data distribution $P_{i}(X,Y)$ for $C_i$. The non-IID data refers to differences between $P_{i}(X,Y)$ and $P_{j}(X,Y)$ for two different client devices $C_i$ and $C_j$. Note that we can factor $P_{i}(X,Y)$ as $P_{i}(Y|X)P_{i}(X)$ or $P_{i}(X|Y)P_{i}(Y)$. Using this factorization, the details of the data generation process is follows.

\begin{itemize}
%\begin{itemize}[leftmargin=*, itemsep=0em, topsep=-5px]
%\setlength\itemsep{0em}
\item {\bf SynCov}: $P_{i}(X)$ varies and $P(Y|X)$ shared among client devices $C_i$. We parameterize $P_{i}(X)$ with a Gaussian distribution $\mathcal{N}(\mu_{i}, \sigma_{i})$ and $P(Y|X)$ with a softmax function, which has weight and bias parameters $W$ and $b$ respectively. First, we sample $W,b \sim \mathcal{N}(0,1)$. Then, for each $C_i$, we sample $\mu_{i}, \sigma_{i} \sim \mathcal{N}(0,1)$ and sample features $x_{i} \sim P_{i}(X)$. We obtain $y_{i}$ from $\argmax_{y \in Y}P(Y|X\,{=}\,x_{i})$.
\item {\bf SynLabel}: $P_{i}(Y)$ varies and $P(Y|X)$ shared among client devices $i$. Given the number classes $|Y|$, we create a discrete multinomial distribution sampled from a Dirichlet distribution Dir$(\beta_{1},\ldots,\beta_{|Y|})$, where $\beta_{k} > 0$. We repeat this for each $C_i$. Then, for each $y \in Y$, we sample $\mu_{y}, \sigma_{y} \sim \mathcal{N}(0,1)$ to parameterize $P(Y|X)$ for all $i$. For each client device we apply logical sampling~\cite{henrion_uai86} to obtain $y_{i} \sim P_{i}(Y)$ and $x_{i} \sim P(X|Y\,{=}\,y_{i})$ accordingly.
\end{itemize}

\textbf{Real-world Datasets}: We evaluate on three standard FL benchmark datasets: MNIST~\cite{726791}, Federated Extended MNIST (FEMNIST)~\cite{cohen2017emnist}, and {\em The Complete Works of William Shakespeare} (Shakespeare)~\cite{mcmahan_fedavg}. These datasets are curated from recently proposed FL benchmarks~\cite{caldas2019leaf,TF-F}. We use the data partition provided by~\cite{li_fedprox}. For MNIST, the data is distributed, via power law, across 1,000 devices and each device has samples of $2$ classes. FEMNIST is a 62 image classification dataset, including both handwritten digits and letters. There are 200 devices and 10 lowercase letters (`a' to `j') is subsampled such that each device has samples for 5 classes. Shakespeare is used for the next character prediction task, consisting of lines spoken by different characters (80 classes) in the plays.

\begin{figure*}[t!]
    \centering
    \subfigure[$\alpha=1$]{
	\includegraphics[width=.3\linewidth]{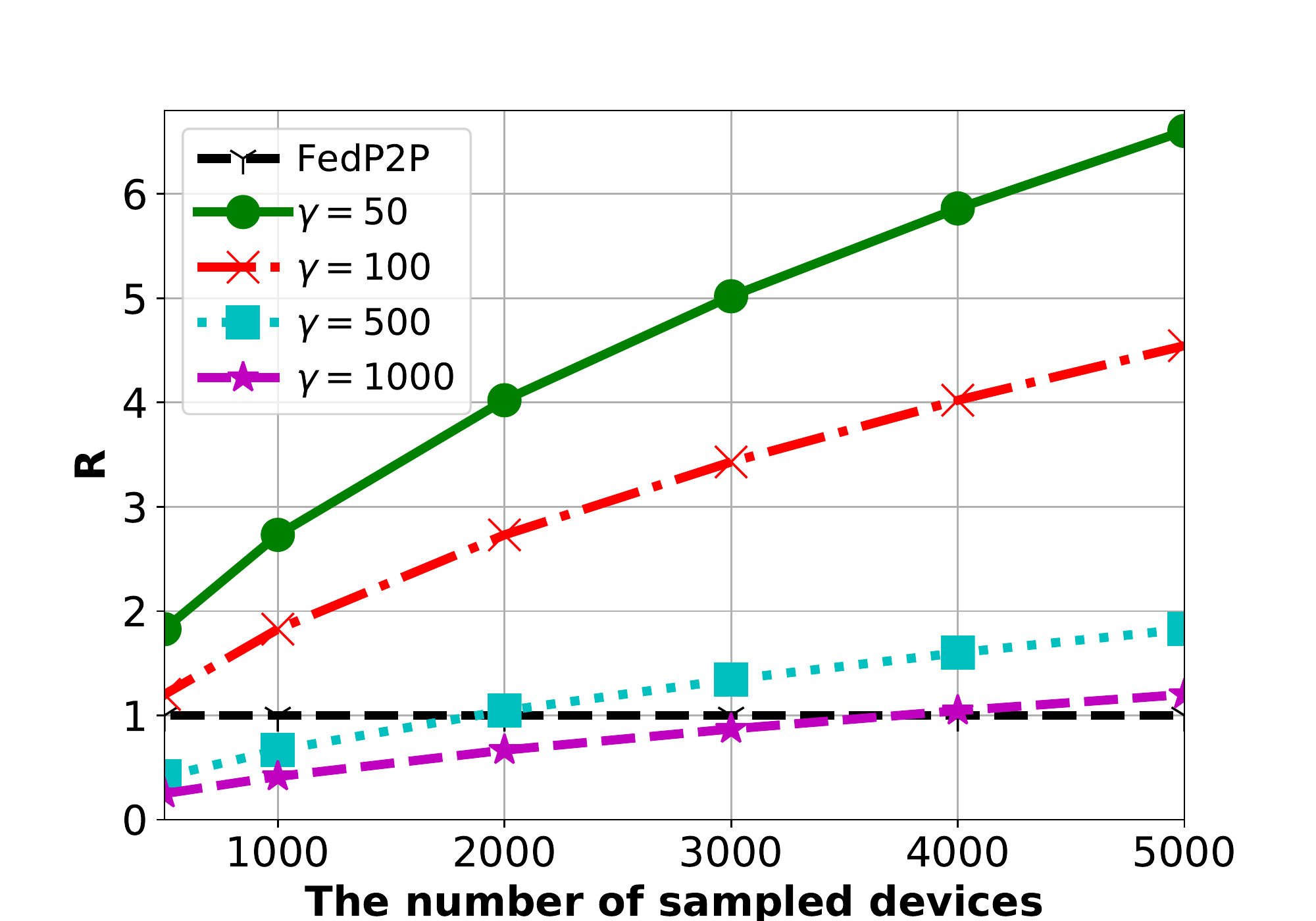}
	\label{fig:fl-sc}
    }    
	\subfigure[$\alpha=4$]{
    \includegraphics[width=.3\linewidth]{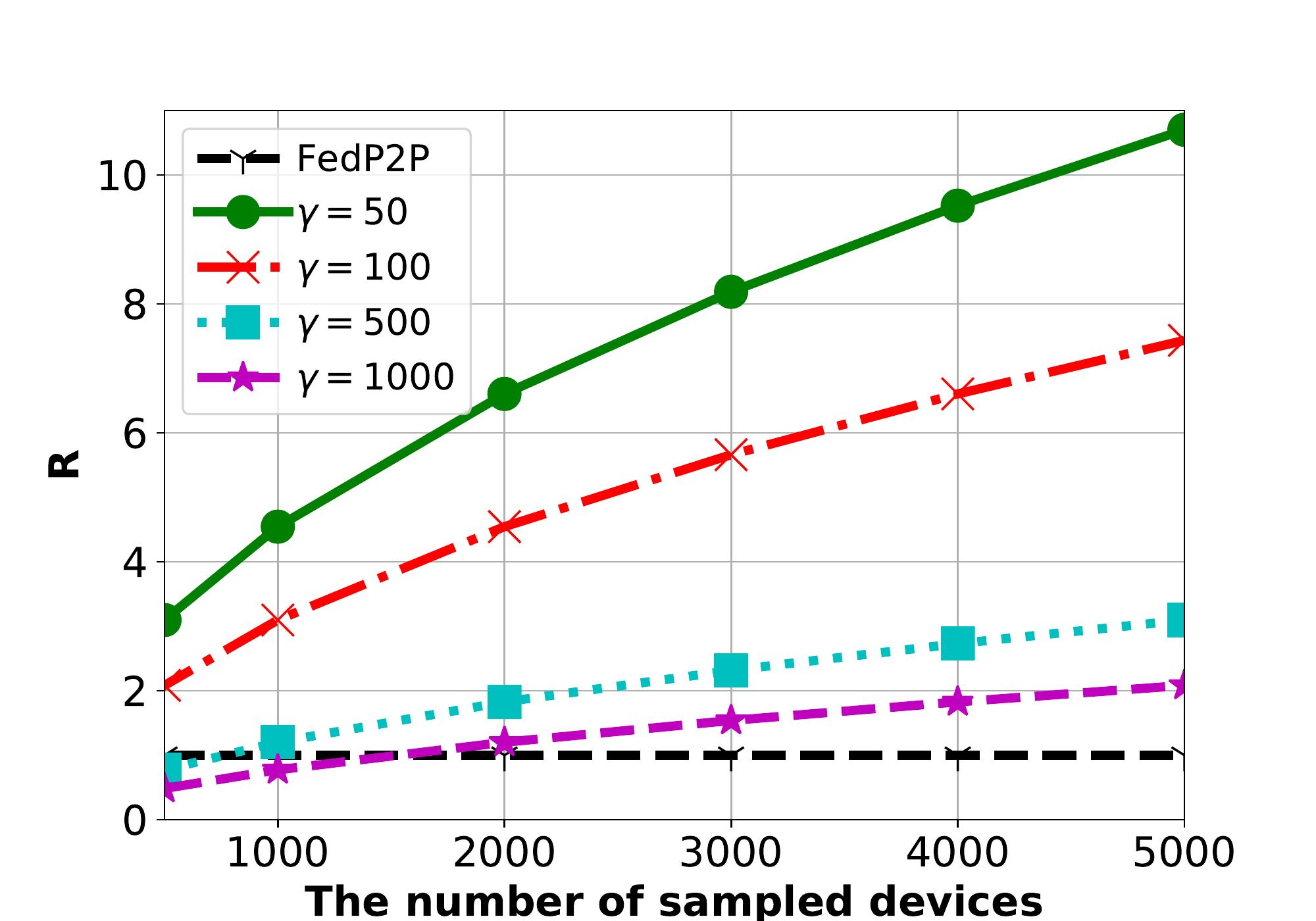}
    } 
	\subfigure[$\alpha=16$]{
    \includegraphics[width=.3\linewidth]{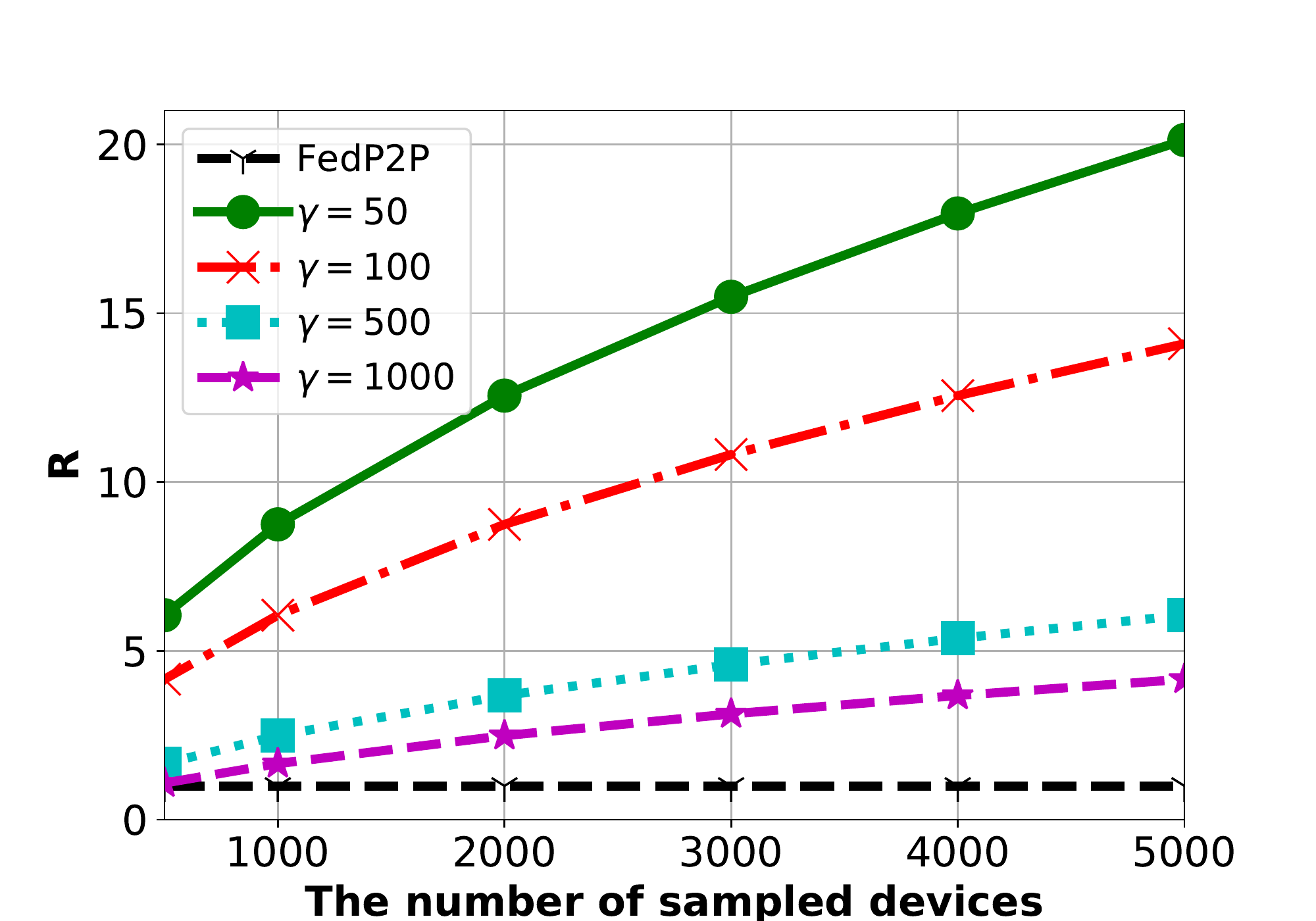}
    }
%    \vspace*{-2mm}
    \caption{Numerical comparison between \fedpp and \fedavg on normalized communication time. Black line is \fedpp and others are \fedavg with various bandwidth settings. \fedavg outperforms \fedpp only when the number of sampled devices is small or the device bandwidth is extremely poor. Detailed discussion is in Section~\ref{sec:excoom}.}
%    \vspace*{-2mm}
    \label{fig:comm} 
\end{figure*}
\begin{figure*}[t!]
    \centering
    \subfigure[SynCov]{
        \includegraphics[trim = 0 0 0 20, clip,width=.3\linewidth]{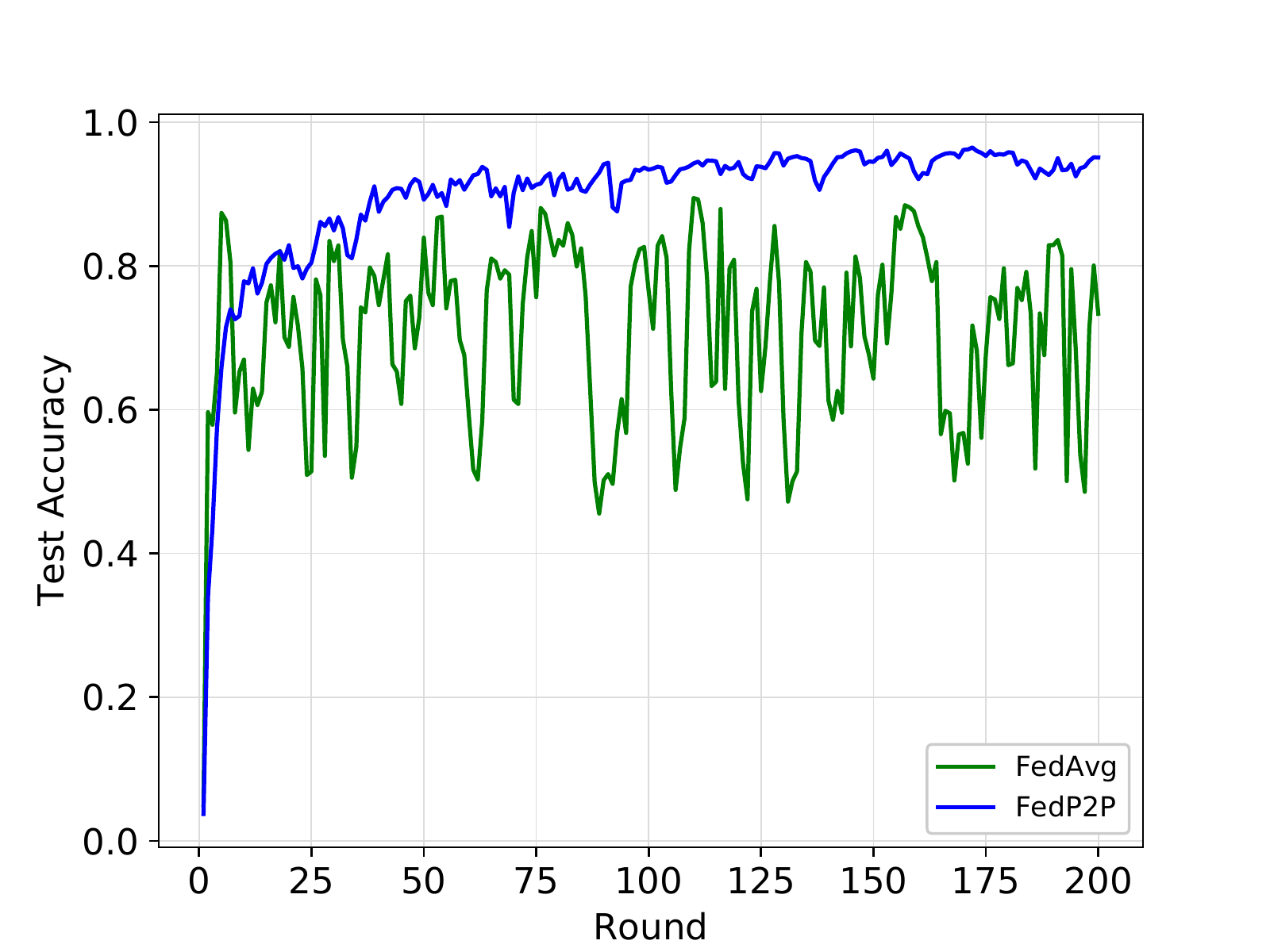}
    }
    \subfigure[SynLabel]{
        \includegraphics[trim = 0 0 0 20, clip,width=.3\linewidth]{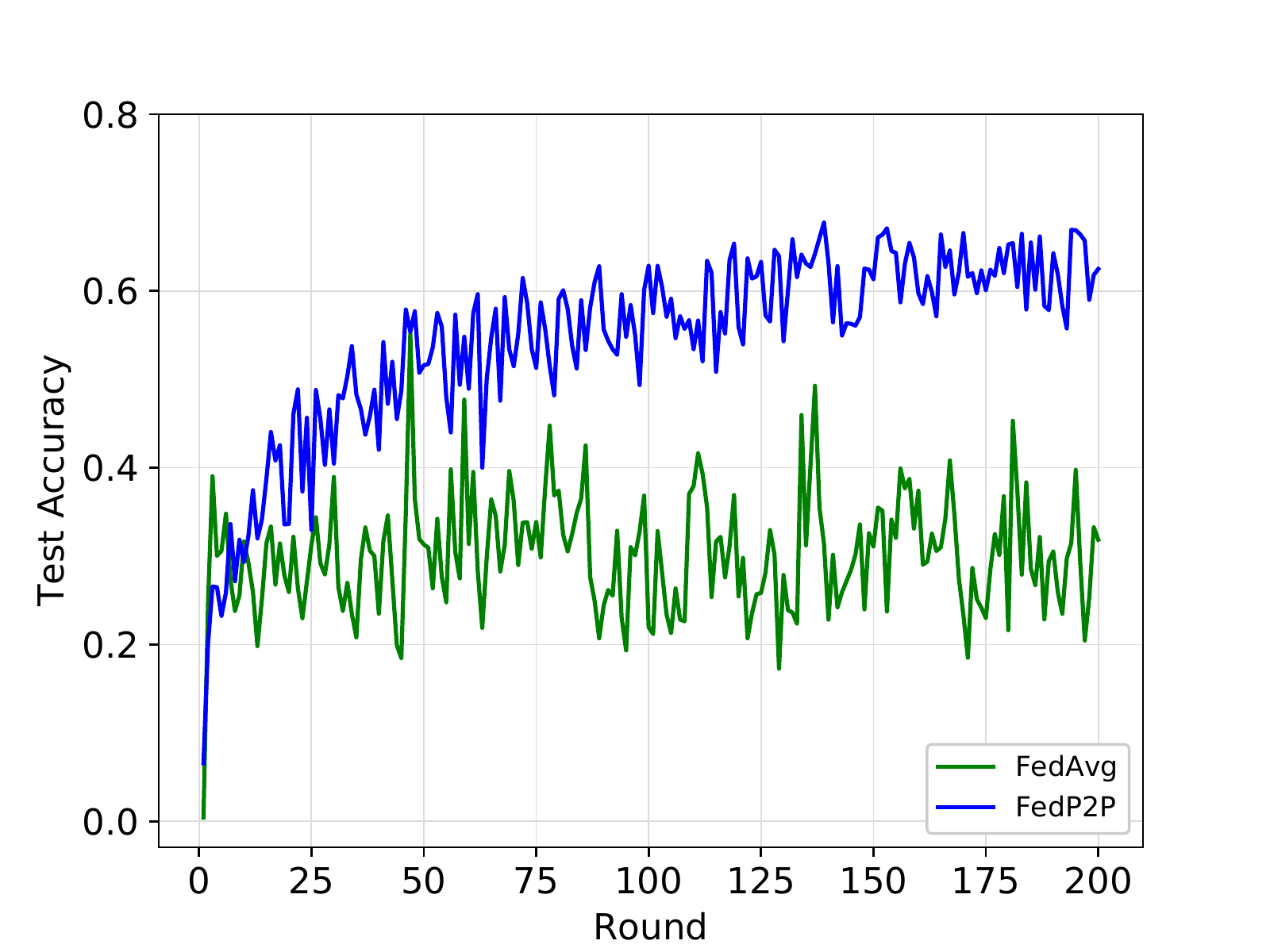}
    }
    \subfigure[MNIST]{
        \includegraphics[trim = 0 0 0 20, clip,width=.3\linewidth]{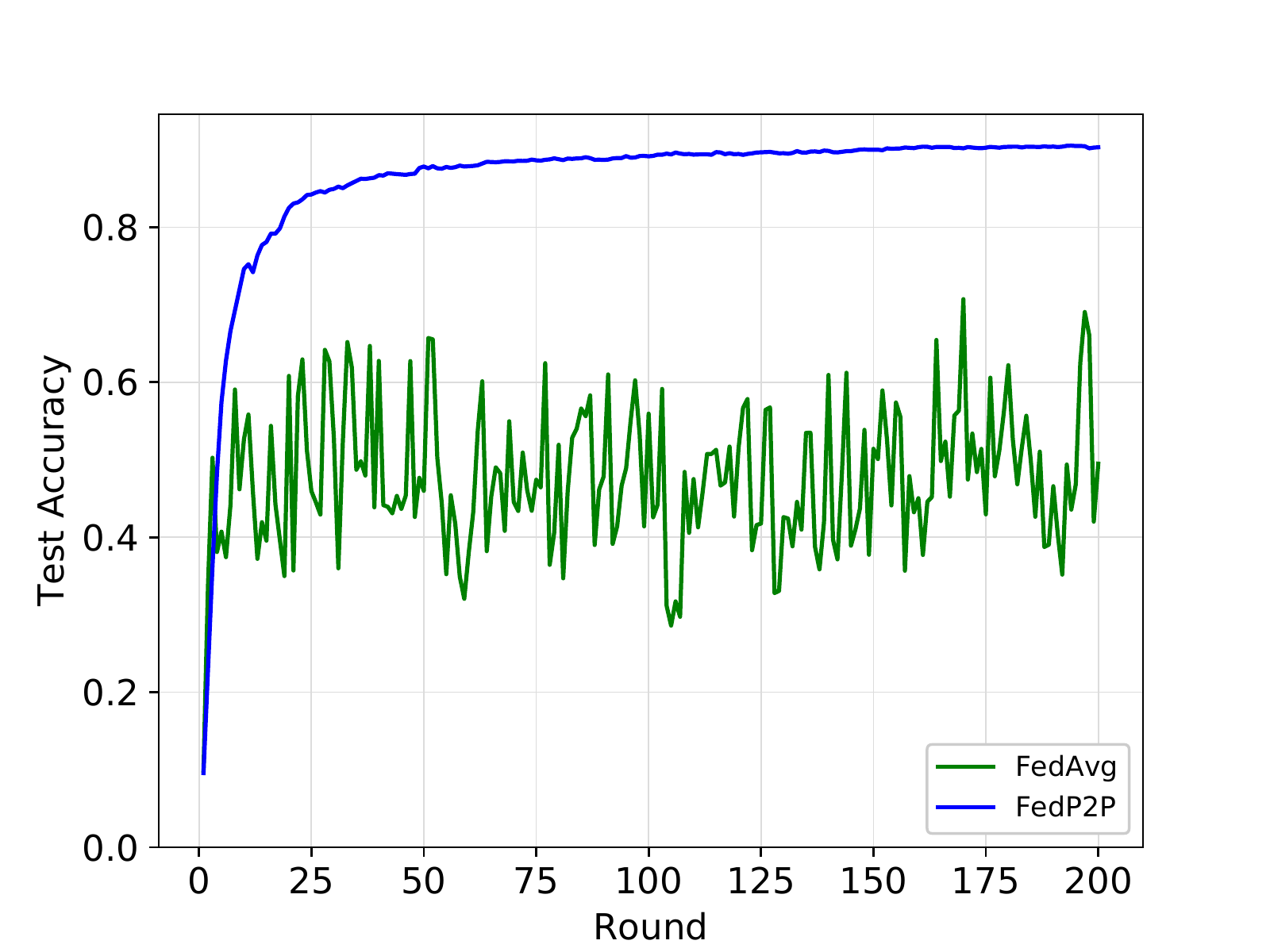}
    }
%    \vspace*{-2mm}
    \caption{Comparison on average test accuracy across clients between \fedavg and \fedpp with 50 \% Stragglers. \fedpp achieves similar accuracy as when there are no stragglers. However, \fedavg suffers from straggler. \fedavg achieves much lower accuracy. Besides, \fedpp shows a much smoother curve. We can observe accuracy jump up to 20 percent between two rounds for \fedavg.}
%    \vspace*{-2mm}
    \label{fig:straggler}
\end{figure*}

\subsection{Implementation}
 We consider centralized FL as our comparison and implemented \fedavg and \fedpp in PyTorch~\cite{pytorch}. We evaluate using various model architectures, both convex and non-convex. We use logistic regression for synthetic and MNIST, Convolution Neural Network for FEMNIST, and LSTM classifier for Shakespeare. In order to draw a fair comparison, we use the same set of models and parameters for \fedavg and \fedpp. Specifically, we use 2-layer CNN with a hidden size of 64 and 1-layer LSTM with a hidden size of 256. Models are trained using SGD. ReLU is used as the activation function. Data is split $80$\% train and $20$\% test. We use batch size of $10$. We use learning rates of $.01$ for synthetic, MNIST, FEMNIST, and and $.5$ for Shakespeare based on grid search on \fedavg and $20$ epochs. Number of devices selected to train per round is fixed to $10$. Both code and datasets will be released once published.

\subsection{Model Accuracy}
In this section, we compare \fedpp and \fedavg in terms of test accuracy. We hold test data for each device and follow the standard evaluation metric of classification accuracy. As shown in Table~\ref{table:real_acc}, \fedpp outperforms \fedavg on all datasets. Fig.~\ref{fig:mainacc} shows the test accuracy as a function of the total round of communication with the central server. Here, we control the number of global communication to be the same for both methods. For a fair comparison, we only let one round of training within each local P2P network. 

On FEMNIST, \fedpp achieved 2.6\% increase in test accuracy. On MNIST, \fedpp achieved 3.7\% increase in test accuracy. On Shakespeare, \fedpp achieved 9\% increase in test accuracy. We see that \fedpp achieves higher testing accuracy on every global communication from Fig.~\ref{fig:mainacc}. Compared to \fedavg, \fedpp gives a rather smooth accuracy curve. In Fig.~\ref{fig:acc-6}, we provide a plot for training loss on FEMNIST. It worth noting that \fedpp also gives a smooth convergence curve. These results demonstrate that \fedpp achieves higher model accuracy compared to the centralized approach. 

\begin{table}
\begin{center}
\caption{Summary of best test accuracy on various datasets for \fedpp and \fedavg.}
%\vspace*{.5em}
\begin{tabular}{ lcc } 
\hline
{\bf Dataset} & \fedpp & \fedavg \\
\hline
MNIST & {\bf0.9092} & 0.8763 \\ 
FEMNIST & {\bf 0.9168} & 0.8929 \\ 
Shakespeare  & {\bf 0.5019} & 0.4563 \\ 
SynCov & {\bf0.9252} & 0.9032 \\ 
SynLabel & {\bf0.6199} & 0.5149 \\ 
\hline
\end{tabular}
%\vspace*{-.5em}
\label{table:real_acc}
\end{center}
\end{table}

\begin{figure*}[t!]
    \centering
    \subfigure[Various choice of $L$]{
	\includegraphics[trim = 100 0 100 20, clip, width=.30\linewidth]{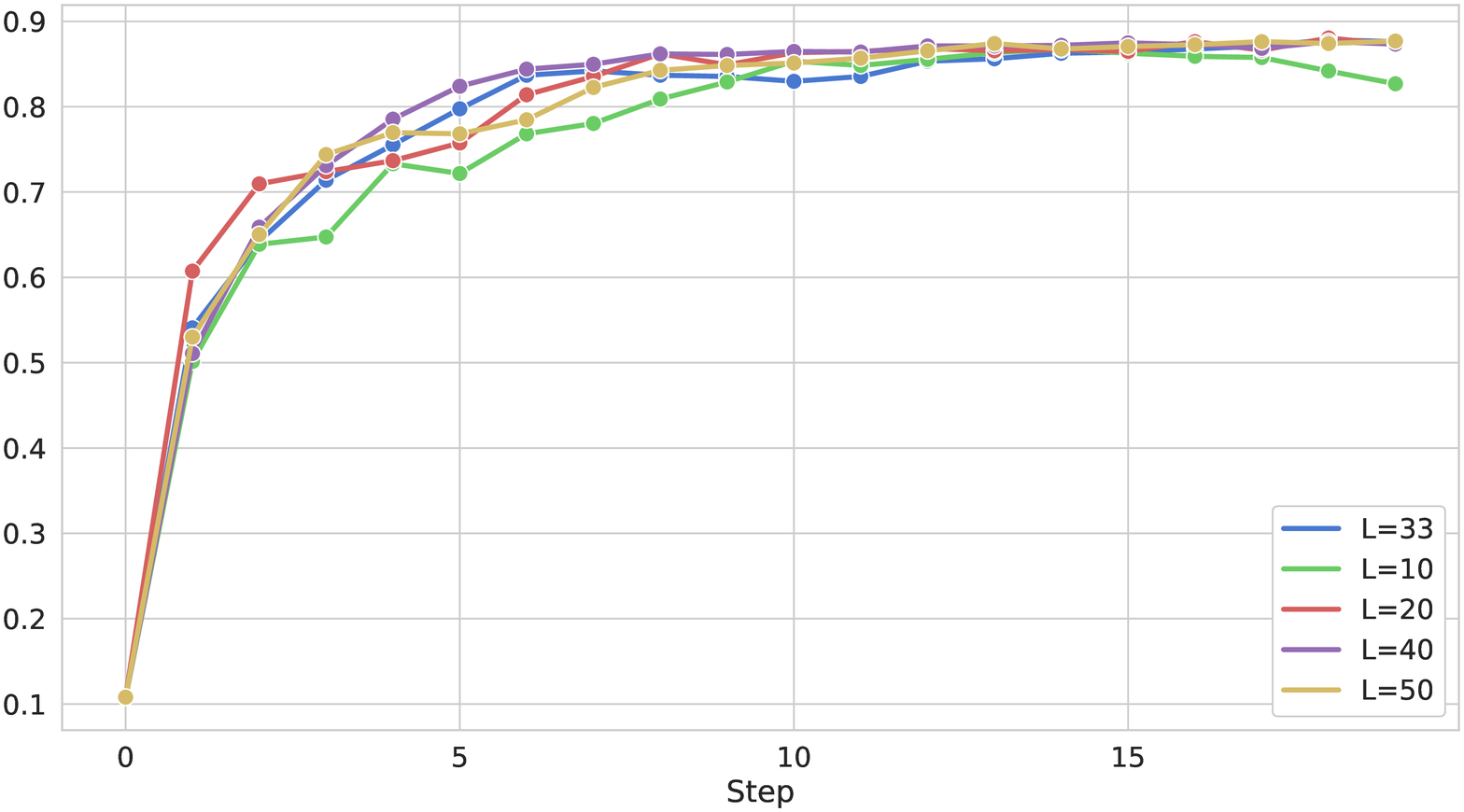}
	\label{fig:ls}
    }
	\subfigure[$P = 100$]{
    \includegraphics[trim = 100 0 100 20, clip,width=.30\linewidth]{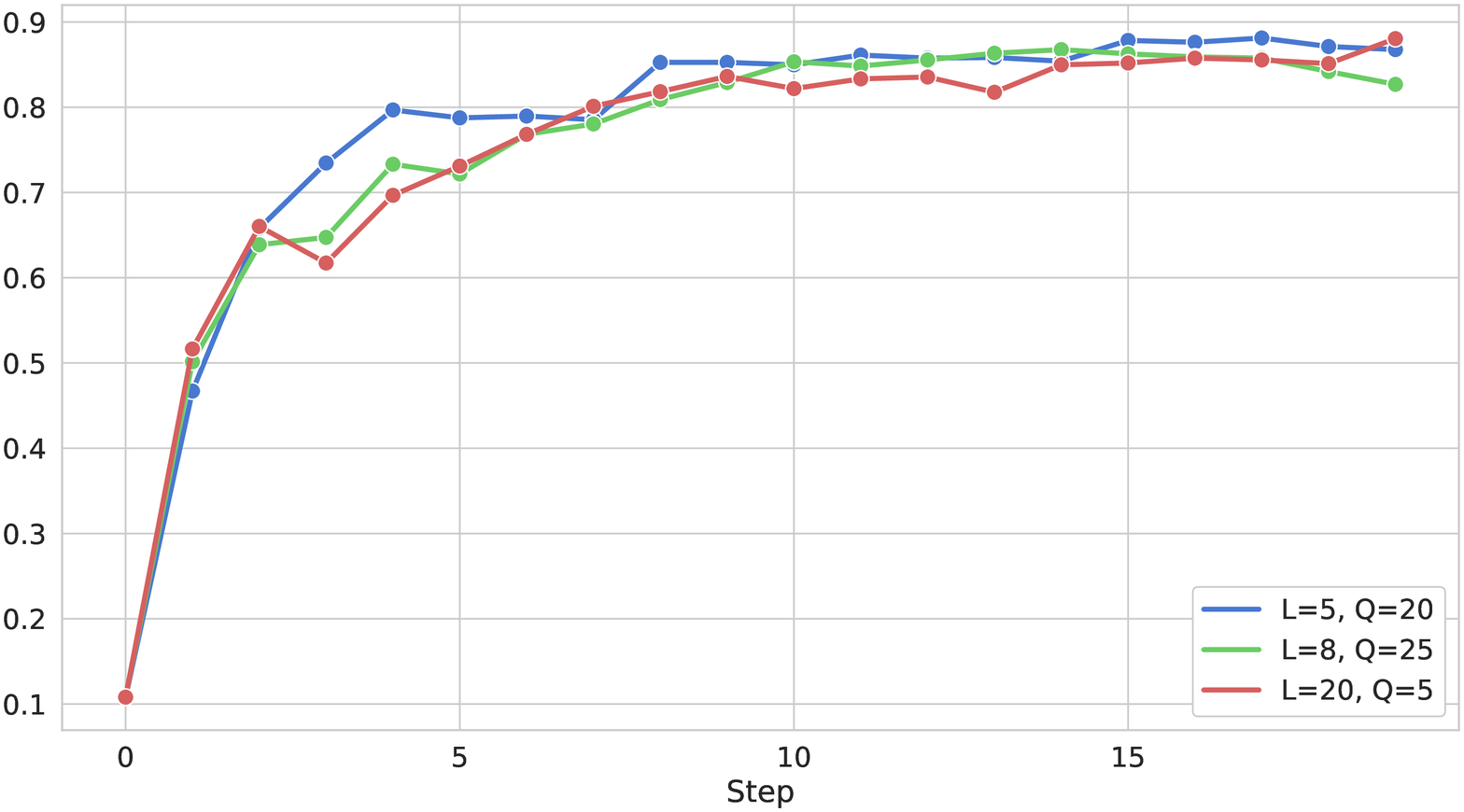}
    \label{fig:lq100}
    } 
	\subfigure[$P = 200$]{
    \includegraphics[trim = 100 0 100 20, clip,width=.30\linewidth]{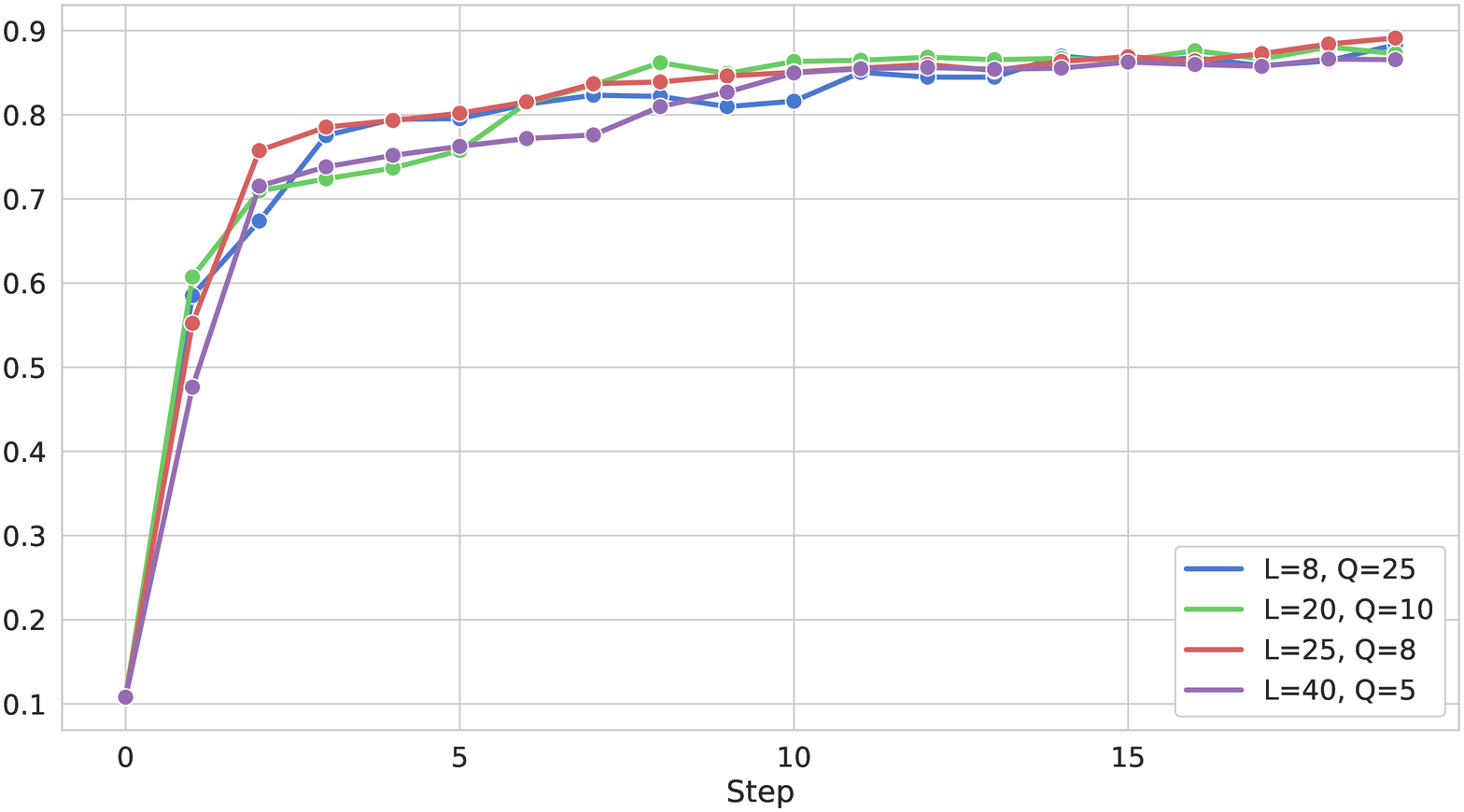}
        \label{fig:lq200}
    }  
    %\vspace*{-3mm}
    \caption{Test accuracy on MNIST for \fedpp on various parameter settings. (a) shows \fedpp with various numbers of local P2P network. Different choices of $L$ do not affect model performance with \fedpp, and thus we can set $L$ to optimize for communication efficiency. (b) and (c) show \fedpp with various number of total participating devices. \fedpp is also robust with diverse choices of $L$ and $Q$.}
        %\vspace*{-3mm}
\end{figure*}

\subsection{Communication Efficiency} 
\label{sec:excoom}
We numerically compare \fedpp and \fedavg in terms of the communication efficiency based upon the analysis in Section \ref{sec:comm}.
The ratio of the download bandwidth to the upload bandwidth of devices varies with different Internet providers, and it covers a wide range.
We set the ratio $\alpha$ to $\{1, 4, 16\}$ as in~\cite{internet_speed}.
The number of sampled devices for one round of training is within $[500, 5000]$~\cite{fl_survey}.
Another parameter that determines the value of $R$ in Eq.~(\ref{eq:comm}) is $\gamma$, which is the ratio of the server bandwidth to the device bandwidth. Current edge devices have the bandwidth of more than 20 Mbps (e.g., 4K streaming~\cite{netflix}). Also, the advent of 5G enables much higher bandwidth for edge devices~\cite{5G_sigcomm}.
The server bandwidth typically ranges from 10 Gbps to less than 1 Gbps~\cite{gaia}.
Therefore, we set $\gamma$ in the range of $[50, 1000]$ in our simulation.

Fig.~\ref{fig:comm} shows the comparison of the communication time.
With $\gamma=100$, \fedpp always outperforms \fedavg with the number of sampled devices larger than 500.
\fedpp can achieve better performance than \fedavg with more sampled devices, smaller $\gamma$, or larger $\alpha$.
However, if the number of sampled devices is small (e.g., $P<100$), or if the device bandwidth is extremely poor (e.g., $\gamma>1000$), then \fedavg can potentially outperform \fedpp in terms of the communication time because the performance bottleneck is not the server in these cases. In practice, thousands of devices are sampled for the training in each round, and the sampled devices usually have high bandwidth due to the sampling mechanism~\cite{fl_survey}. For example, the server tends to select devices with powerful computing capacity and high bandwidth to avoid stragglers. As a result, we argue that \fedpp is more scalable and communication-efficient than \fedavg for large-scale FL.

\subsection{Stragglers Effect and Choice of $\boldsymbol{L}$ and $\boldsymbol{Q}$}

{\bf Straggers}: It is known that FL suffers from stragglers. Specifically, devices fail in on-device training or drop connection due to hardware issues or communication instability. Stragglers cause convergence problems and lead to models with poor performance. We empirically test how robust \fedpp is in the straggler situation in terms of model performance, and we present our result in Fig.~\ref{fig:straggler}. We drop $50$\% of the selected devices to simulate stragglers. \fedpp performs exceptionally well with stragglers. \fedpp archives similar accuracy while the performance of \fedavg dramatically drops. Besides, comparing to \fedavg, accuracy curves are still relatively smooth for \fedpp. For example, the most significant jump on MNIST for \fedavg can exceed over 20 percent, while we do not observe a noticeable increase with \fedpp. 

{\bf $\boldsymbol{L}$ and $\boldsymbol{Q}$}: We conduct two sets of experiments on MNIST to illustrate how \fedpp performs under various parameters. \fedpp introduces $L$ local P2P networks, and within each P2P network, $Q$ devices participated in the training. We conduct the first set of experiments with varying $L$ and the same $Q$ and present the result in Fig.~\ref{fig:ls}. We do not observe significant differences among various $L$. Compared with \fedavg from Fig.~\ref{fig:acc-1}, we notice that all \fedpp plots lies above \fedavg. Thus, we conclude that \fedpp is robust to various $L$ setting from the perspective of convergence and accuracy. As a result, in practice, \fedpp allows us to choose $L$ to optimize for communication efficiency. We conduct the second set of experiment with various combination of $L$ and $Q$, where $P = L \times Q$ and $P = 100$ and $200$. We present the result the in Fig.~\ref{fig:lq100} and Fig.~\ref{fig:lq200}. We observe that different combinations of $L$ and $Q$ has negligible effect on classification performance.

\section{Conclusion}

In this work, we focus on decentralizing FL and propose \fedpp, an approach that utilizes peer communication to train one global model collectively. A unique possibility due to our random partition process is that it allows us to exploit device network topology. If we assume that the data distribution is independent of the network evolving topology, then a random selection of devices to form local P2P networks is identical with any deterministic selection (i.e., Principle of deferred decisions~\cite{mitzenmacher_book}). Effectively, we can partition devices into local P2P networks based on favorable network topology. For example, it is widely accepted that long communication hops create potential problems such as low throughput and high latency due to network congestion. Grouping devices based on communication hops would greatly benefit communication efficiency. For example, in Fig.~\ref{fig:fedp2p-graph}, devices within fewer communication hops are grouped into the same local P2P network.

%In this work, we focus on decentralizing federated learning and propose \fedpp, a federated learning approach that utilizes peer communication to train one global model collectively. \fedpp achieves communication efficiency by distributing communication workload to edge devices and possibly incorporating network topology information.  Our partition and aggregation design yield high-performance models due to variance reduction and allows for the incorporation of network topology information to even further optimize communication. Our empirical evaluation across a range of benchmark datasets provides strong evidence to support our claim and demonstrate that \fedpp enjoys benefits from both communication and model efficiency.

%
%
%
%
% the environments 'definition', 'lemma', 'proposition', 'corollary',
% 'remark', and 'example' are defined in the LLNCS documentclass as well.

\bibliographystyle{splncs04}
\bibliography{fedp2p_bib}

\end{document}